\documentclass[twocolumn]{aastex631}

\usepackage{graphicx}% Include figure files
\usepackage[caption=false]{subfig}
\usepackage{dcolumn}% Align table columns on decimal point
\usepackage{bm}% bold math
\usepackage{xspace}
\usepackage{pifont}
\usepackage{braket}
\usepackage{hyperref}

\usepackage{siunitx}

\graphicspath{{.}}
\newcommand{\AEI}{Max-Planck-Institut für Gravitationsphysik, Albert-Einstein-Institut, Callinstr.~38, D-30167 Hannover, Germany}
\newcommand{\LUH}{Leibniz Universität Hannover, D-30167 Hannover, Germany}
\newcommand{\CITA}{Canadian Institute for Theoretical Astrophysics, 60 St George St, University of Toronto, Toronto, ON M5S 3H8, Canada}
\newcommand{\DADDAA}{David A. Dunlap Department of Astronomy and Astrophysics, University of Toronto, 50 St. George St., Toronto ON. M5S 3H4, Canada}
\newcommand{\UofTPhys}{Department of Physics, University of Toronto, 60 St. George St., Toronto, ON M5S 3H8, Canada}

\shorttitle{Spin Distribution of Retained Black Holes from Hierarchical Mergers}
\shortauthors{Borchers, Ye \& Fishbach}

\begin{document}

\title{Gravitational-wave kicks impact spins of black holes from hierarchical mergers}

\author[0000-0002-2184-7388]{Angela Borchers}
\affiliation{\AEI}
\affiliation{\LUH}
\author[0000-0001-9582-881X]{Claire S.\ Ye}
\affiliation{\CITA}
\author[0000-0002-1980-5293]{Maya Fishbach}
\affiliation{\CITA}
\affiliation{\DADDAA}
\affiliation{\UofTPhys}

\date{\today}

\begin{abstract}	
One proposed black hole formation channel involves hierarchical mergers, where black holes form through repeated binary mergers. Previous studies have shown that such black holes follow a near-universal spin distribution centered around 0.7. However, gravitational-wave kicks can eject remnants from their host environments, meaning only retained black holes can participate in subsequent mergers. We calculate the spin distribution of retained black holes in typical globular clusters, accounting for remnant kick velocities. Since the kick magnitude depends on the binary's mass ratio and spin orientations, certain configurations are more likely to be retained than others. This preferentially selects certain remnant spin magnitudes, skewing the spin distribution of second-generation black holes away from the universal distribution. In low escape velocity environments, the distribution can become bimodal, as remnants with spins of 0.7 typically receive larger kicks than other configurations. Regarding higher-generation black holes, their spin distribution does not converge to a unique form, and can span a broad range of spins, $a_f \in (0.4,1)$, depending on their merger history, birth spins and the escape velocity. Additionally, we find that the presence of a small fraction of binaries with near-aligned spins can produce a second, more dominant peak, whose position depends on the birth spin magnitude. Our findings identify observable features of hierarchical merger black holes, which is essential for understanding their contribution to the gravitational-wave population. Moreover, the dependence of the spin distribution on astrophysical parameters means that precise spin measurements could provide insights into their formation environments.
\end{abstract}

\section{Introduction}

Black holes formed through hierarchical mergers of lighter black holes are expected to have a unique spin distribution, peaking around $a_f \sim 0.7$ \citep{Berti:2008af, Fishbach:2017dwv}. Such mergers typically occur in dense environments where black holes are dynamically assembled into binaries, leading to isotropic spin orientations \citep{PortegiesZwart:1999nm, Rodriguez:2016vmx, Gerosa:2021mno}. These distinctive features are often used to characterise binary black hole mergers from dynamical origins and differentiate them from other formation channels, such as isolated field formation, where black holes are expected to have small, aligned spins \citep{Kalogera:1999tq, Bavera:2020inc, Gerosa:2018wbw, Callister:2020vyz}. 
	
Because black-hole spins leave a specific imprint in the gravitational-wave (GW) signal, it is possible to estimate them from GW observations \citep[see e.g.,][]{Chatziioannou:2024hju}. Current GW observations generally provide poor constraints on black hole spins for individual events \citep{KAGRA:2021vkt, PhysRevD.89.064048, Chatziioannou:2014coa, Purrer:2015nkh}, as these measurements are limited by a partial degeneracy between the mass ratio and a combination of the spin parameters \citep{Cutler:1994ys, PhysRevD.52.848, Baird:2012cu, Ohme:2013nsa, Hannam:2013uu}. However, improved detector sensitivity will enable more precise spin measurements in the future.

Despite the uncertainties in the spin estimates of individual events, after three observing runs of the GW detector network \citep{KAGRA:2021vkt} formed by Advanced LIGO \citep{LIGOScientific:2014pky}, Advanced Virgo \citep{VIRGO:2014yos} and KAGRA \citep{KAGRA:2020tym}, it is possible to analyse the spin distribution of the binary black hole population \citep{KAGRA:2021duu}. The spin and mass distributions can be used to infer the formation origin of the population of black holes \citep[e.g.][]{Stevenson:2015bqa, Rodriguez:2016vmx, Fishbach:2017dwv, PhysRevD.95.124046, Farr:2017uvj, Farr:2017gtv, Fishbach:2018edt, Gerosa:2021hsc, Fishbach:2021yvy, Farah:2023vsc, Callister:2023tgi, Banagiri:2025dxo, KAGRA:2021duu}. To do so, it is crucial to have accurate models of the expected parameter distributions in different formation channels. While early works by \cite{Tichy:2008du, Lousto:2009ka} derived statistical distributions of spin magnitudes for generic binary configurations, \cite{Fishbach:2017dwv} investigated the spin distribution specific to hierarchical mergers. In the hierarchical merger scenario, the spin distribution has been shown to be unique, approximately independent of the mass distribution (assuming only a preference for major mergers with near-unity mass ratios), the black hole birth spins, and the merger generation \citep{Fishbach:2017dwv}.

In this paper, we revisit the problem of determining the spin distribution of black holes from hierarchical mergers, incorporating a key consideration: the GW remnant kick velocities produced in binary black-hole mergers. Remnant kicks occur due to the anisotropic emission of linear momentum through GWs during a merger. Because of momentum conservation, any loss of linear momentum is balanced by a recoil kick imparted to the merger remnant \citep{1961RSPSA.265..109B, Bekenstein:1973zz, Fitchett:1983qzq}. The magnitude of this kick depends on the properties of the parent black holes, and it can be as large as $5000\,{\rm km\,s^{-1}}$ for certain binary configurations \citep{Gonzalez:2007hi, Bruegmann:2007bri, Campanelli:2007cga, Lousto:2011kp, Lousto:2019lyf}. These velocities often exceed the escape velocities of dense star clusters, potentially ejecting merger remnants from their host environments. If a black hole is ejected, it is unlikely it will find a companion and merge again in the galactic field. Thus, for a GW signal to originate from a binary involving a black hole formed in a previous merger, that black hole must have remained within its environment before acquiring a new companion. Therefore, in this study, we compute the kick velocities of binary mergers and determine the spin distribution of \textit{retained} remnants, particularly in dense star clusters.

Although several studies have highlighted the role of kicks in hierarchical mergers \citep[e.g.][]{Merritt:2004xa, Gerosa:2019zmo, Doctor:2021qfn, Mahapatra:2021hme, Mahapatra:2022ngs, Alvarez:2024dpd}, the spin distribution of retained black holes remains largely unexplored. \cite{Alvarez:2024dpd, Mahapatra:2024qsy} propose methods to assess whether the black holes from individual GW events originate from hierarchical mergers by calculating both the final spin and the kick velocity. In particular, \cite{Alvarez:2024dpd} explore how the relation between these two quantities changes with different initial spin configurations. Yet, none of these studies have explicitly derived the spin distribution of retained black holes.

For some time, many studies argued that the birth spin of black holes from stellar collapse was highly uncertain~\citep[see e.g.][]{Miller:2014aaa, Kushnir:2016zee}. While some recent studies suggest that black-hole birth spins are generally small \citep{Fuller:2019sxi, Bavera:2020inc}, GW observations indicate that black holes have non-zero spins, reaching up to 0.4 \citep{Callister:2022qwb}. In this work, we do not restrict ourselves to zero birth spins, but instead, we explore how the birth spin magnitude impacts the spin distribution of retained black holes across multiple merger generations. Furthermore, it has traditionally been argued that black holes in dense star clusters have isotropic spin orientations~\citep[e.g.][]{Rodriguez:2016vmx}, but there may be alternate mechanisms to align some binaries' spins~\citep[e.g.][]{Kiroglu:2025bbp}. We investigate whether a small fraction of binaries with nearly aligned spin can impact the spin distribution of black holes from hierarchical mergers.

\section{Methods}

We calculate the spin distribution of black holes formed through hierarchical mergers. Specifically, we are interested in black holes that are retained by their host environments, which depends on the kick velocities of the merger remnant. The kick velocity of a remnant black hole is determined by General Relativity and depends on the initial properties of the binary. 

Approximately $10\%$ of binaries in dense star clusters are expected to have non-zero orbital eccentricities \citep{Rodriguez:2017pec}. While post-Newtonian estimates suggest that the kick scales as $1 + e$ \citep{Sopuerta:2006et}, recent studies indicate that eccentricity can amplify the kick magnitude by up to 25\%, making it harder for eccentric binaries to be retained in such environments \citep{Sperhake:2019wwo, Radia:2021hjs}. Despite these findings, the phenomenology of kicks in eccentric binaries remains insufficiently explored, and no fitting formula currently exists to estimate the kicks from generic binaries. Therefore, in our study, we assume binaries have quasi-circular orbits. Given the small fraction of eccentric binaries in clusters and their high likelihood of ejection, this assumption is unlikely to significantly impact our results.

Given that the kick is determined by the mass ratio and the initial spins, only certain combinations of masses and spins will result in retained remnants. At the same time, the spin of the remnant black hole is determined by the same initial properties. Intuitively, it can be understood as the sum of the initial spins and the angular momentum of a test particle at the last stable orbit around a Kerr black hole \citep{Buonanno:2007sv}. Therefore, the parameter combinations that lead to the retention of a particular black hole will also determine its final spin value.

\subsection{Predicting the black hole spin distribution}

\begin{figure}[tb]
	\includegraphics[width=0.45\textwidth]{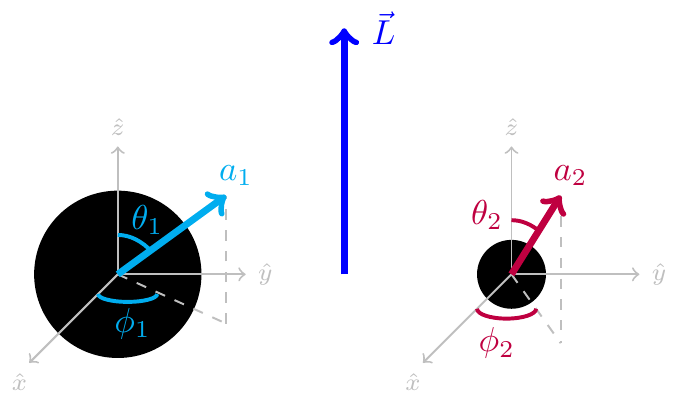}
	\caption{The spin parameters of the primary and secondary black holes. It is common to describe the spins by their magnitudes $a_1$, $a_2$, their tilt angles $\theta_1$, $\theta_2$, and their azimuthal angles $\phi_1$, $\phi_2$, in a reference frame where the $z$-axis is aligned with the orbital angular momentum of the system, $\vec{L}$.}
	\label{spin_parameters}
\end{figure}

We consider a large ensemble of black holes, $N = 10^6$, and randomly select pairs for mergers. First-generation black holes are assumed to have the same spin magnitude, $a_i$, while higher-generation black holes inherit their spins from previous mergers. The spin orientations of all binary black holes are random and isotropic. Figure \ref{spin_parameters} illustrates the corresponding spin parameters. For each pair, we compute the final spin and the kick velocity using the state-of-the-art fitting formula for black-hole remnant properties, \texttt{NRSur7dq4Remnant} \citep{Varma:2018aht}. In our model, only black holes with kick velocities smaller than the escape velocity of typical globular clusters can undergo subsequent mergers; those with higher kicks are removed from the ensemble.

We refer to black holes born from stellar collapse as first-generation black holes (1G), while we refer to those black holes that are remnants from a black-hole binary merger as second-generation black holes (2G) or, in a more general case, $n$-th generation black holes.

Since hierarchical mergers typically occur in dense star clusters, we also consider the characteristic mass ratio distribution found in these environments. In particular, we use data from the \texttt{Cluster Monte Carlo} (CMC) cluster catalog, which includes simulations of globular clusters with different initial properties \citep{Kremer:2019iul}. Our analysis incorporates data from all 144 CMC simulations currently available. 

\begin{figure}[tb]
	\includegraphics[width=0.45\textwidth]{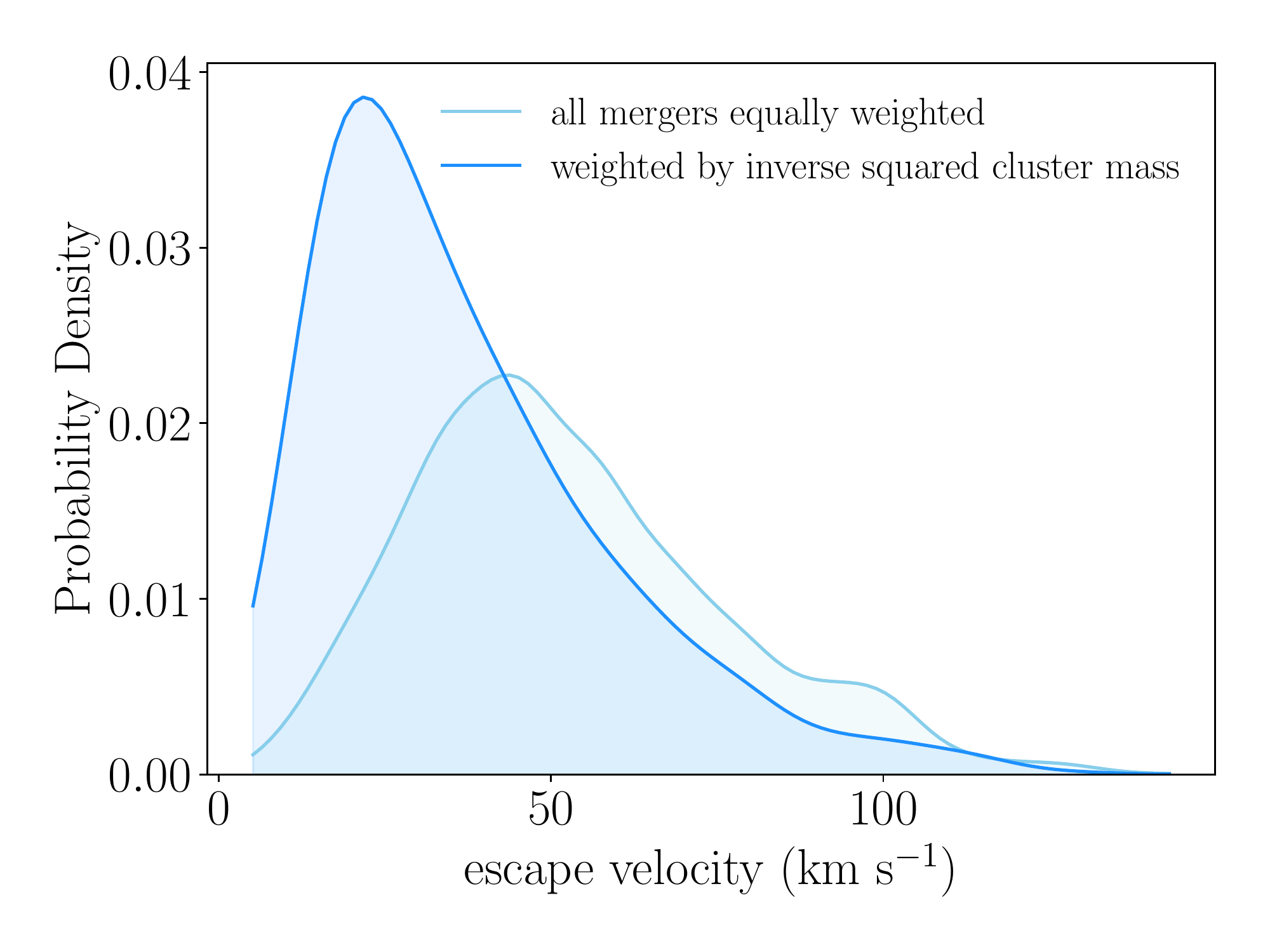}
	\caption{Escape velocity profile of all 1G+1G mergers in CMC simulations. The light blue distribution represents escape velocities with equal weighting for all mergers, while the dark blue distribution is weighted by host cluster's mass ($\propto M^{-2}$). These profiles do not reflect the escape velocity distribution of a single cluster but rather that of all CMC clusters, each weighted differently. In our analysis, we use the equal-weight distribution, as it is a more conservative assumption. It peaks at approximately $50\,{\rm km\,s^{-1}}$ and decreases to zero around $5\,{\rm km\,s^{-1}}$ and $145\,{\rm km\,s^{-1}}$ respectively.}
	\label{escape_velocity_profile_CMC}
\end{figure}

To determine the spin distribution of remnants from 1G+1G mergers, we use escape velocity data from all mergers across CMC simulations. Since the cluster mass function is uncertain, the escape velocity distribution of clusters is also uncertain. Here we consider two possible escape velocity distributions, each weighted differently. Figure \ref{escape_velocity_profile_CMC} illustrates these distributions: the light blue curve considers equal weights for all mergers, while the dark blue curve weights them by the host cluster's mass, following a $\propto M^{-2}$ scaling. As we show in Sec.~\ref{sec:impact_of_escape_velocity}, lower escape velocities reduce retention rates and amplify the impact of kicks on the spin distribution of retained black holes. Therefore, among the two distributions, the equal-weight distribution represents a more conservative choice, as it peaks at higher escape velocities. In our analysis, we adopt this distribution, which peaks at approximately $50\,{\rm km\,s^{-1}}$ and decreases to zero around $5\,{\rm km\,s^{-1}}$ and $145\,{\rm km\,s^{-1}}$, respectively. 

It is important to note that this profile does not represent the escape velocity distribution of a single globular cluster, but rather a combination of multiple clusters with varying masses and radii. For each black hole pair in our model, we randomly sample from this escape velocity distribution. We then compare the binary's kick velocity to the sampled escape velocity to decide whether the remnant is retained or ejected. When analyzing the convergence of the spin distribution across multiple merger generations, we assume a fixed escape velocity and test a range of values, $v_{\text{esc}} \in (50, 100, 200)\,{\rm km\,s^{-1}}$. 

\begin{figure}[tb]
	\includegraphics[width=0.45\textwidth]{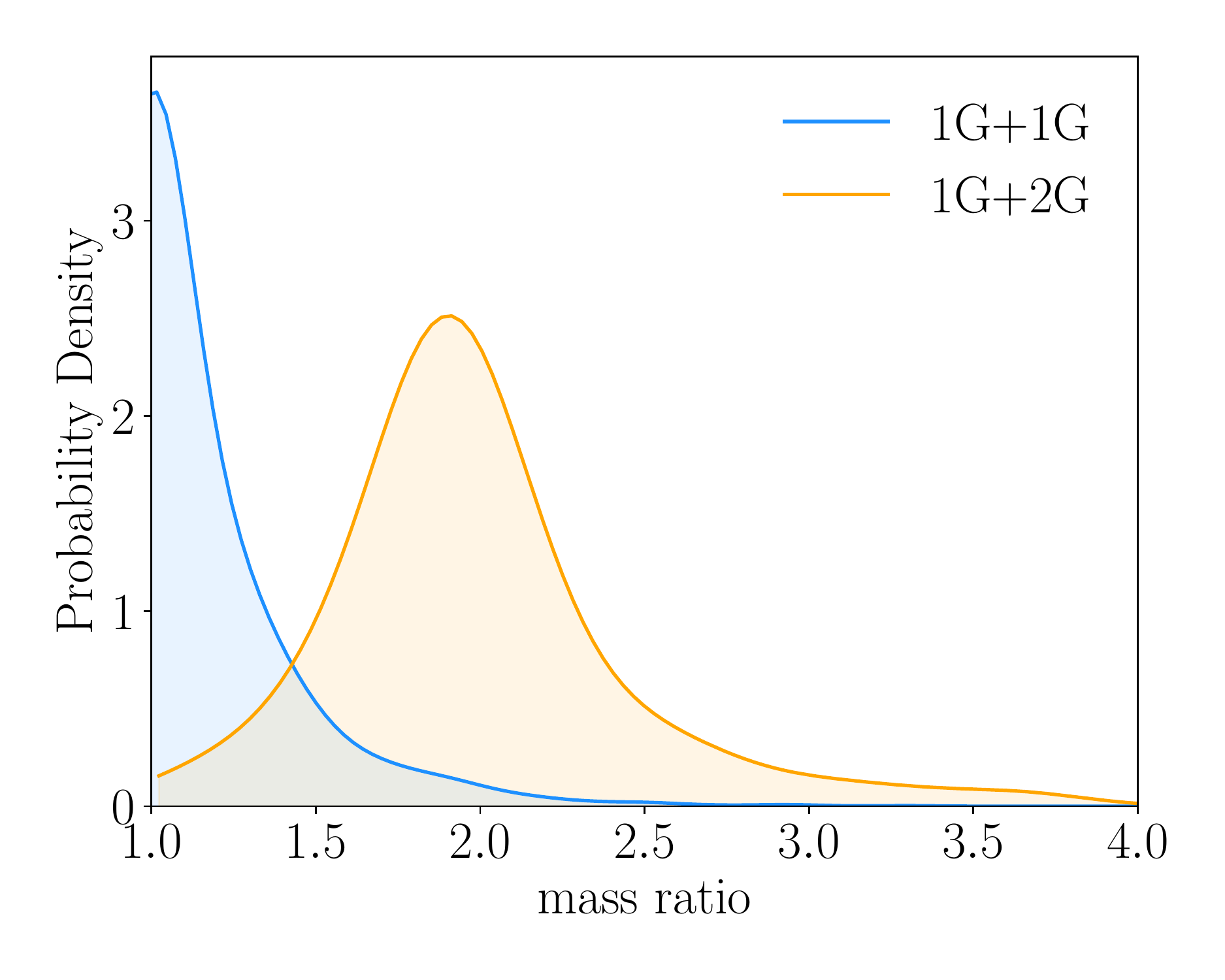}
	\caption{Mass ratio distributions of 1G+1G (blue) and 1G+2G (orange) mergers obtained from CMC simulation data, equally weighting all mergers.}
	\label{mass_ratio_distributions}
\end{figure}

We extract the mass ratio distributions of 1G+1G and 1G+2G black-hole mergers from CMC simulations, weighting all mergers equally. Figure \ref{mass_ratio_distributions} presents these distributions. Mass ratios are reported as the ratio of the primary (more massive) black hole to the secondary (less massive) black hole, so that mass ratios are greater than 1. For the rest of the higher-generation mergers (i.e. 2G+2G and above), the CMC data lacks sufficient samples to produce a reliable distribution. In these cases, we adopt a fixed mass ratio value. Motivated by the values at which the mass ratio distributions of 1G+1G and 1G+2G mergers peak, we define the mass ratio as $m/n$, where $m$  and $n$ denote the generations of the primary and secondary black holes, respectively.

\subsection{The kick velocity across the binary black hole parameter space}
Even though we consider isotropically oriented spins, the remnant kick velocity is sensitive to the masses and especially the spin orientations of black hole binaries \citep[see e.g.,][]{Gonzalez:2006md, Herrmann:2007ac, Koppitz:2007ev}. This means that certain binary configurations have higher chances of being retained than others. 

\begin{figure}[tb]
	\includegraphics[width=0.45\textwidth]{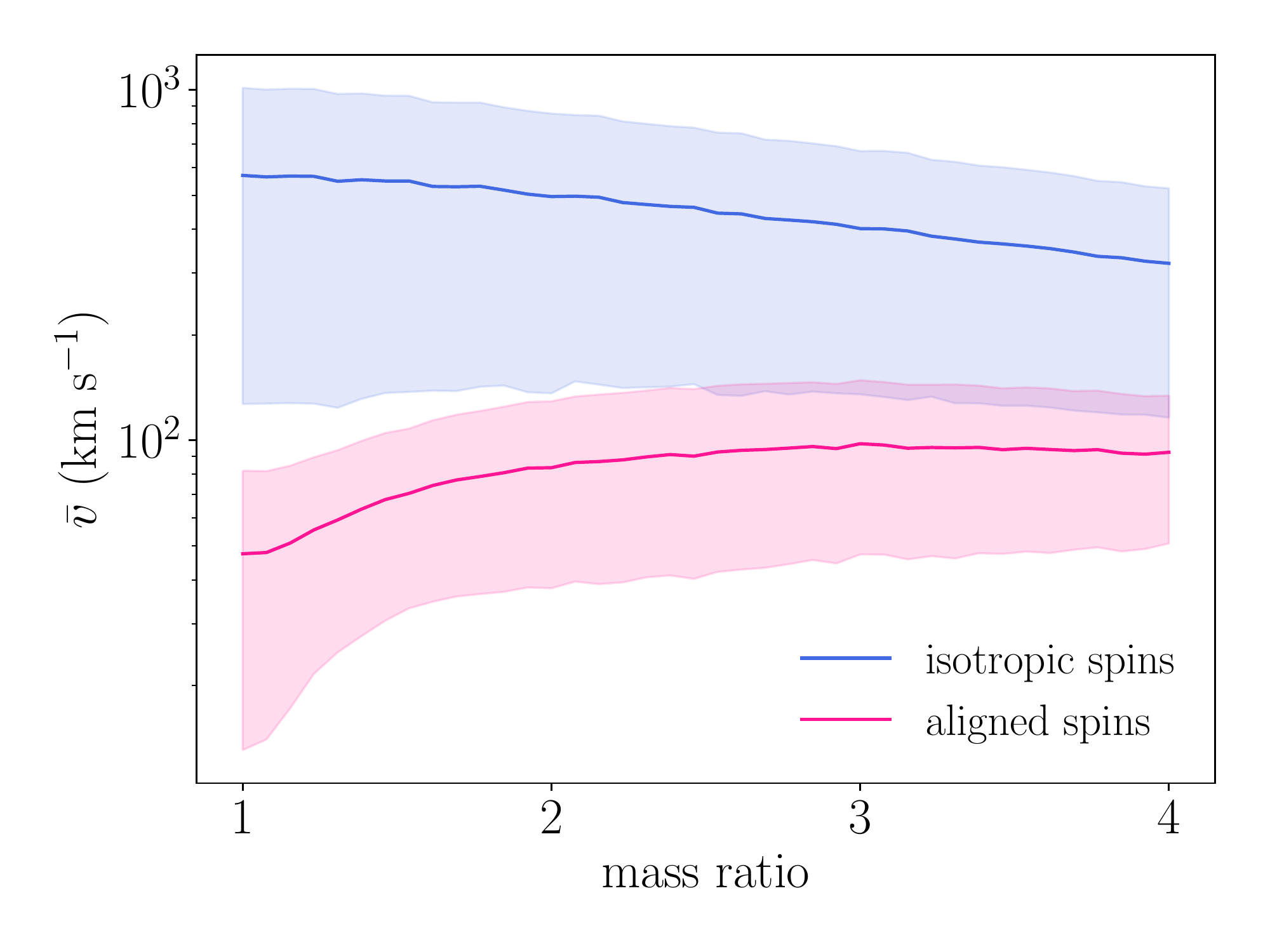}
	\caption{The kick velocity dependence on the mass ratio for two cases: binaries with random isotropic spins (blue) and binaries with aligned spins (pink). For each mass ratio value, we have generated $10^4$ binary configurations with spins sampled uniformly in the following way. In the isotropic case, for each black hole, the spin magnitude $a_i$ is drawn from $[0, 0.8]$, the spin tilt angle from $\cos\theta_i \in [-1, 1]$ and the spin azimuthal angle $\phi_i$ from $[0, 2\pi]$. For aligned spins, we sample the spin magnitude of each black hole uniformly from $[0, 0.8]$. The solid line represents the mean kick velocity, while the shaded region indicates one standard deviation. The kick estimates were computed using the \texttt{NRSur7dq4Remnant} fitting formula.}
	\label{mass_ratio_dependency_of_the_kick}
\end{figure}

\begin{figure*}[tb]
        \centering
        \subfloat[$q = 1$]{%
		\includegraphics[width=0.33\linewidth]{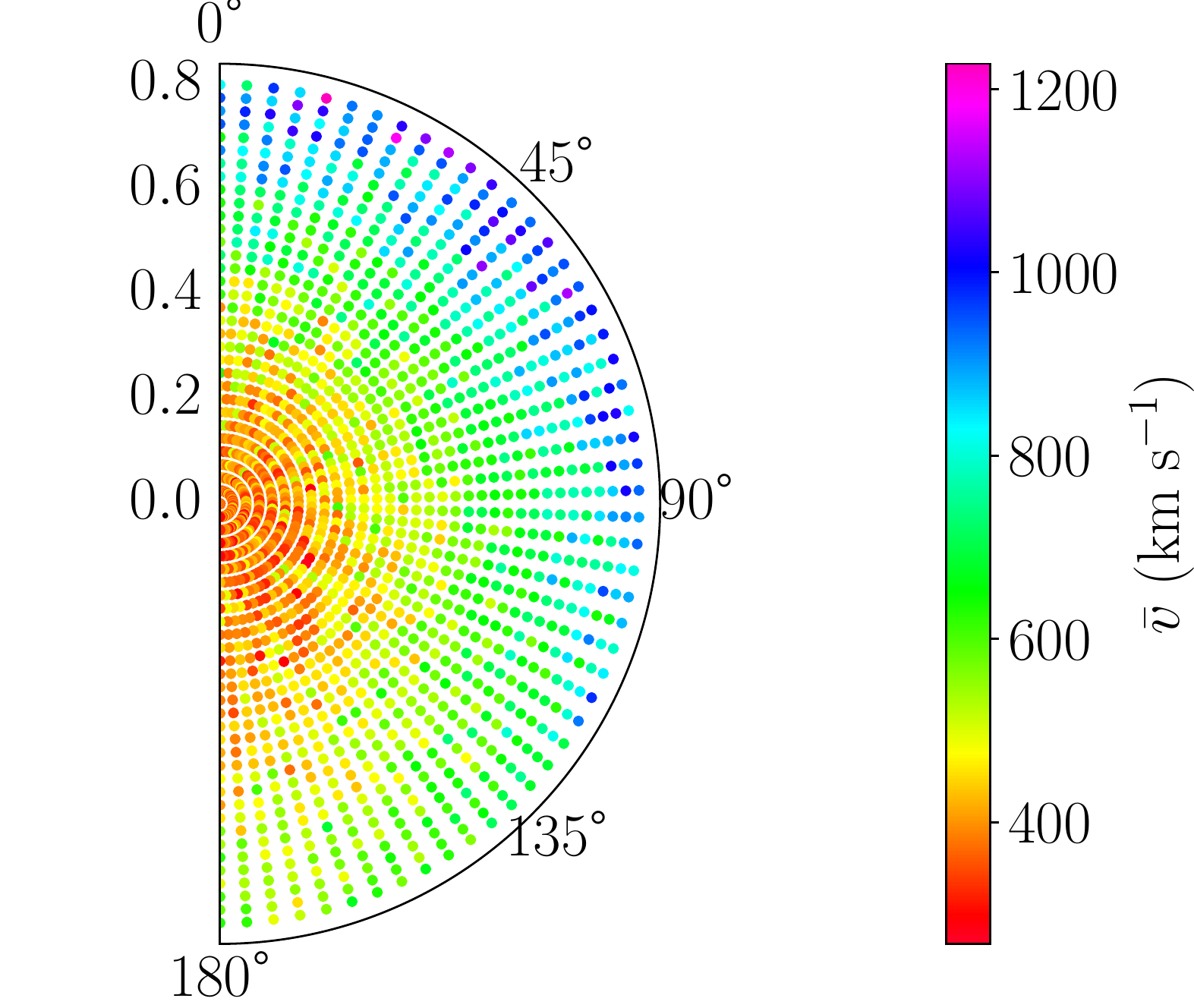}}
	\subfloat[$q = 2$]{%
		\includegraphics[width=0.33\linewidth]{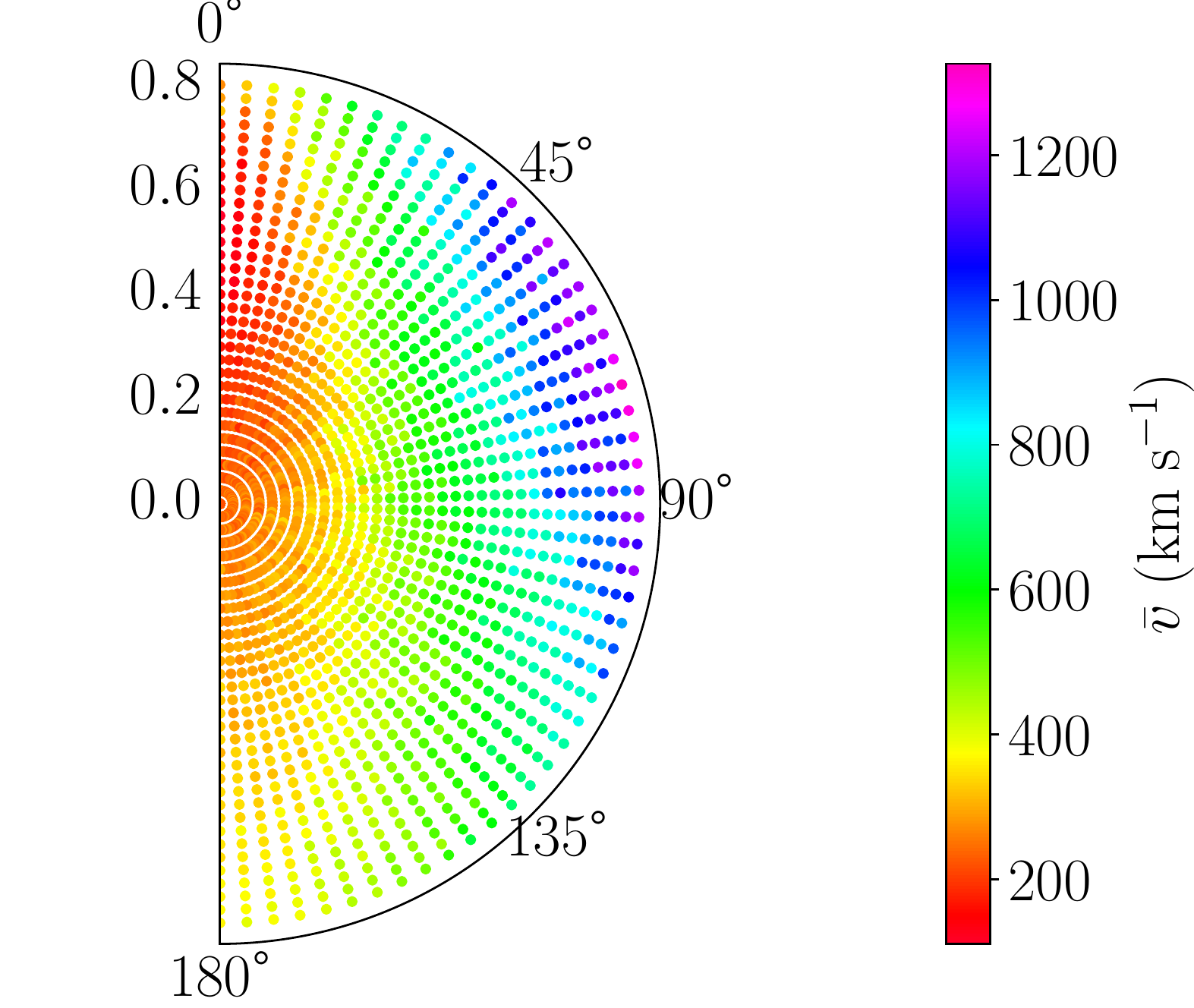}}
        \subfloat[$q = 4$]{%
		\includegraphics[width=0.33\linewidth]{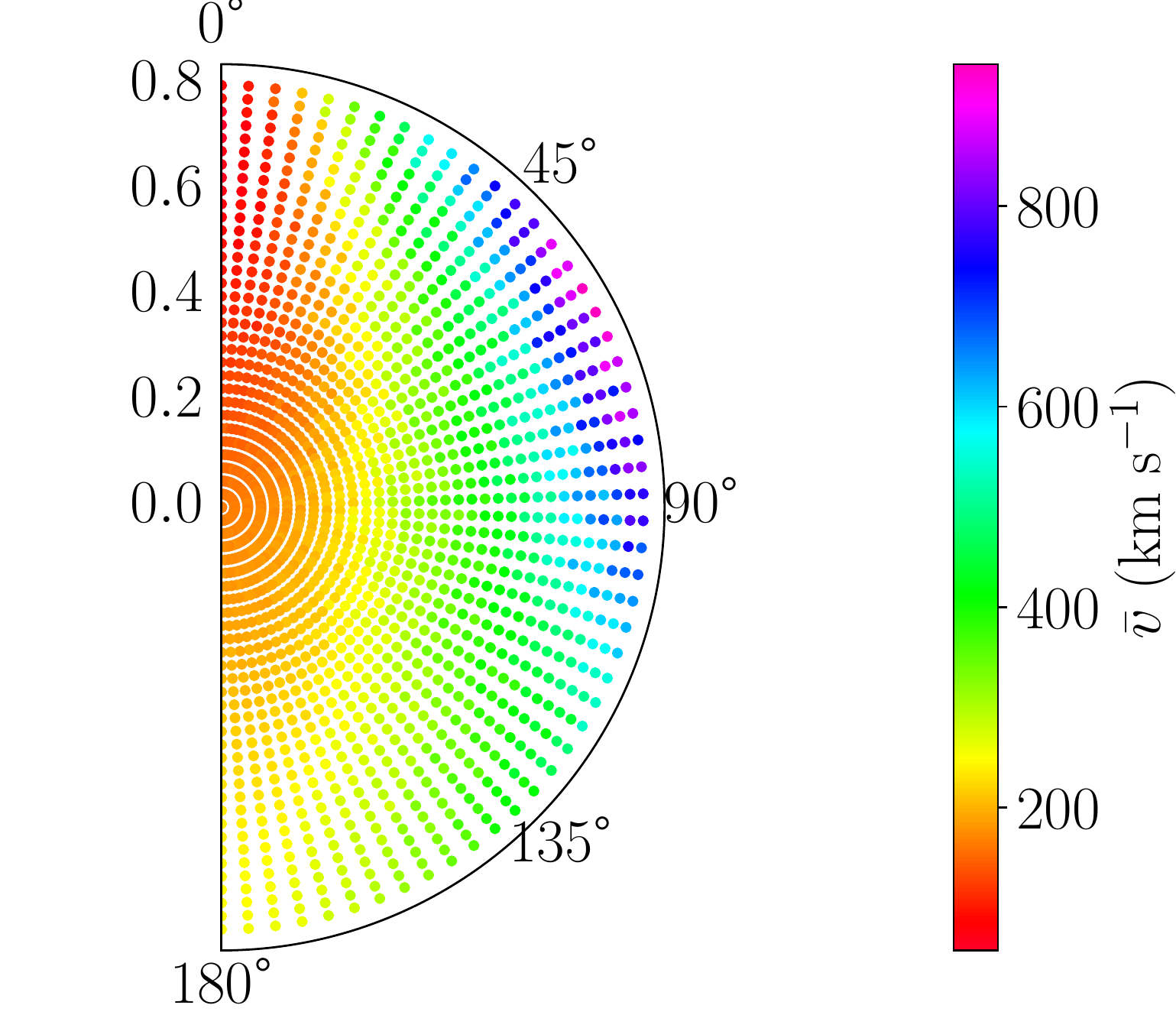}}
	\caption{Kick velocities of double-spin binaries with mass ratios $q = 1$, $q = 2$ and $q =4$ respectively. In each polar plot, every dot represents a different spin configuration. The radial distance indicates the primary spin magnitude $a_1$, while the tilt angle of the polar plot represents the primary spin tilt angle $\theta_1$. For every primary spin configuration, we compute the kick velocity, choosing a random secondary spin magnitude and orientation. The points in the polar plots are colored according to the kick velocity. The kick magnitude shown is averaged over 50 different configurations, where the primary spin magnitude and tilt are fixed (specified by the position of the dot in the plot), and the primary spin azimuthal angle and the secondary spin are randomly chosen. Kick velocities have been computed using the \texttt{NRSur7dq4Remnant} fit.}
	\label{kicks_from_double-spin_binaries}
\end{figure*}

In binaries where both black holes are spinning, isotropic spins produce significantly larger kicks than aligned spins, as illustrated in Figure \ref{mass_ratio_dependency_of_the_kick}. Specifically, for a mass ratio of $q = 1$, isotropic spins yield an average kick magnitude of $570 \pm 443\,{\rm km\,s^{-1}}$, whereas aligned spins produce an average kick of $47 \pm 34\,{\rm km\,s^{-1}}$, with the error bars representing 1$\sigma$ deviations from the mean values. Figure \ref{kicks_from_double-spin_binaries} shows in more detail how different spin orientations influence the resulting kick velocity, assuming different values of the binary's mass ratio. The position of each colored dot indicates the spin magnitude and spin tilt of the primary black hole, while the color indicates the kick velocity averaged over 50 different binaries with a random, isotropic secondary spin direction and magnitude, and a random primary spin azimuthal angle. As we increase the spin magnitudes, the plot shows that aligned-spin binaries have smaller kicks than binaries with in-plane or anti-aligned spins. The difference between the kick estimates of these configurations becomes more pronounced as we increase the mass ratio to $q = 4$. In \autoref{sec:appendix}, we investigate in more detail the reason why aligned-aligned binaries have the smallest kicks in unequal-mass binaries. These results suggest that retained remnants have higher chances of having parents with aligned spins. 

Regarding the mass ratio dependency of the kick, equal-mass systems have, on average, larger kick magnitudes than unequal-mass systems when assuming isotropic spins, as illustrated by Figure \ref{mass_ratio_dependency_of_the_kick}. Therefore, their remnants have higher chances of being ejected than those from unequal-mass binaries. On the other hand, the kick velocities from aligned-spin binaries do not depend on the mass ratio so strongly.

\subsection{Impact of black-hole spins on the retention rate}

As the kick velocity is sensitive to the spin magnitudes and orientations, we investigate how the retention rate of 1G+1G mergers depends on the black-hole spins. We consider an ensemble of $10^6$ black holes and explore five cases: (i) both black holes have random isotropic spins (representing a 1G+1G or 2G+2G merger) (ii) the primary black hole has a random isotropic spin and the secondary is nonspinning (representing a 1G+2G merger) (iii) the primary black hole has an aligned spin, and the secondary is nonspinning, (iv) the primary is anti-aligned and the secondary is aligned, (v) and both black holes have aligned spins. In the case where both black holes have spin, we assume both black holes have the same spin magnitude. For this calculation, we use the mass ratio distribution and escape velocity profile of 1G+1G mergers from CMC data (see Figures \ref{mass_ratio_distributions} and \ref{escape_velocity_profile_CMC}).

\begin{figure*}[tb]
\centering
	\includegraphics[width=0.85\textwidth]{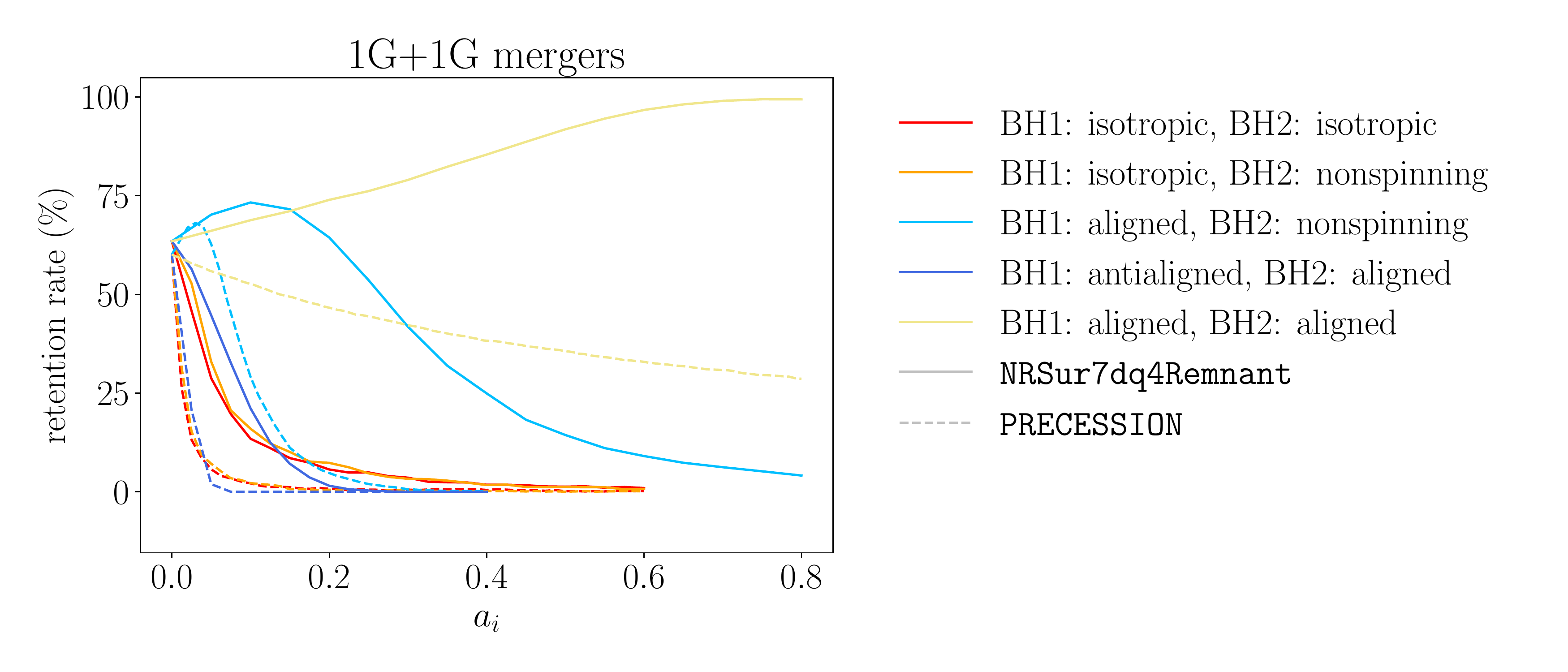}
	\caption{Retention rate of 1G+1G mergers as a function of the initial spin magnitude assuming an escape velocity profile from CMC data, averaged over cluster properties. The plot shows how the retention rate changes depending on the spin orientations considered for the components of the binary. In red, assuming both black holes have random isotropic spins; in orange, assuming the primary has a random isotropic spins, and the secondary is nonspinning; in light blue, assuming the primary has an aligned spin and the secondary is nonspinning; in dark blue, assuming the primary has an anti-aligned spin and the secondary an aligned spin; and in yellow, assuming both black holes have aligned spins. Solid lines show the estimates using \texttt{NRSur7dq4Remnant}, while dashed lines show the estimates using \texttt{PRECESSION}. In the legend, we use the acronym ``BH" to refer to black hole.}
	\label{retention_rates}
\end{figure*}

Figure \ref{retention_rates} shows the retention rate as a function of the initial spin magnitude for the five cases we mentioned. We observe that the retention rate decreases with increasing spin magnitude in most of the cases, except for the aligned-nonspinning case, where the retention rate increases up to initial spins of $a_i = 0.2$ before decreasing to zero. Interestingly, the retention rate of aligned-aligned binaries increases with the initial spin magnitude. This is because the kick magnitude decreases with the increase in spin magnitudes (see \autoref{sec:appendix} for more details). On the other hand, the retention rate of isotropic spin binaries decreases faster than for other spin configurations, as these binaries produce larger kicks than aligned spin binaries.

Many of the codes for simulating dense star clusters, including the CMC code, use the fitting formula from \cite{Gerosa:2016sys} to estimate the final spin and the kick velocity. This prescription is available through the Python package \texttt{PRECESSION}. Since the development of this model, the catalog of numerical relativity simulations has increased in size and accuracy \citep[see e.g.,][]{Boyle:2019kee, Healy:2022wdn, Hamilton:2023qkv}, allowing the development of more accurate remnant models in the last years \citep{Jimenez-Forteza:2016oae, Varma:2018aht, Boschini:2023ryi, Planas:2024vnq}. As we mentioned earlier, in our study, we use the state-of-the-art model \texttt{NRSur7dq4Remnant} from \cite{Varma:2018aht}. To understand whether modelling differences can impact kick estimates and, in turn, retention rates, we compare the estimates of \texttt{PRECESSION} and \texttt{NRSur7dq4Remnant}. 

\begin{figure}[tb]
	\includegraphics[width=0.45\textwidth]{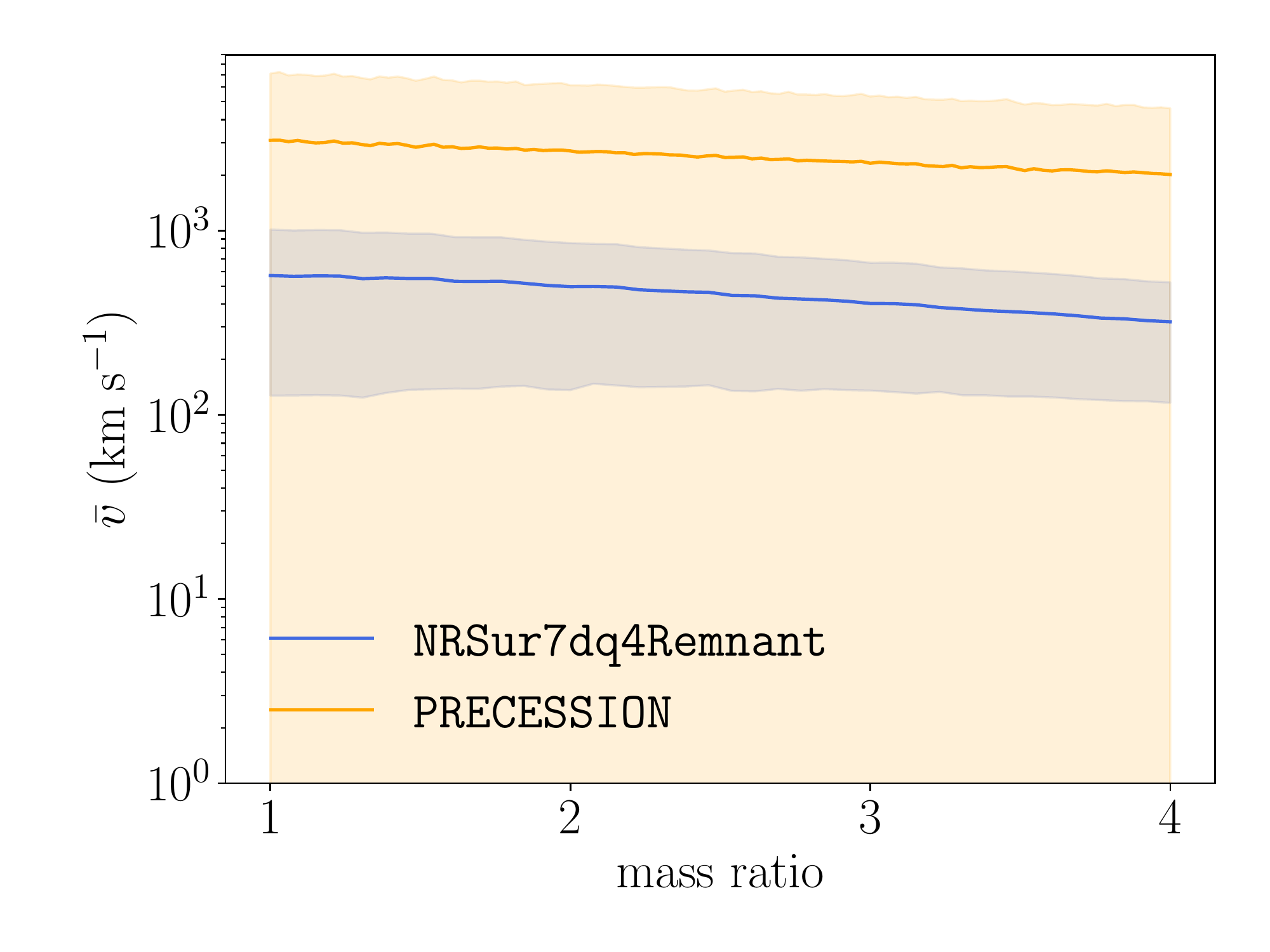}
	\caption{Kick estimates of \texttt{NRSur7dq4Remnant} (blue) and \texttt{PRECESSION} (orange) for $2 \times 10^6$ binary configurations with random spin magnitude and spin orientations as a function of the mass ratio. Solid lines indicate the mean value, while the shaded region indicates one standard deviation from the mean value. Given the improved numerical relativity calibration of the model, \texttt{NRSur7dq4Remnant} is more accurate than \texttt{PRECESSION}. As indicated by the solid lines, on average, \texttt{NRSur7dq4Remnant} predicts kick values that are an order of magnitude smaller than those from \texttt{PRECESSION}, and its estimates have a significantly smaller spread.}
	\label{kick_estimates_model_comparison}
\end{figure}

We find that systematic errors can impact retention rate estimates when 1G black holes have non-zero spins. The estimates differ even for small initial spins of $a_i = 0.1$. As noted by \cite{Gerosa:2016sys}, \texttt{PRECESSION} loses accuracy in the equal-mass limit and when dealing with nonspinning and single-spin binaries. Figure \ref{kick_estimates_model_comparison} shows the differences between the kick estimates of \texttt{NRSur7dq4Remnant} and \texttt{PRECESSION} for $2 \times 10^6$ different binary configurations with random spin magnitudes and orientations as a function of the mass ratio. We have sampled the spin magnitude uniformly over $a_i \in [0, 0.8]$, the spin tilt angle uniformly in $\cos\theta_i \in [-1, 1]$ and the spin azimuthal angle over $\phi_i \in [0, 2\pi]$. As Figure \ref{kick_estimates_model_comparison} shows, on average, \texttt{NRSur7dq4Remnant} predicts kick magnitudes that are an order of magnitude smaller than those from \texttt{PRECESSION} and has a significantly smaller spread in the kick estimates. Finding disagreements between different models is not uncommon, as kick estimates are particularly sensitive to waveform systematic errors \citep{Borchers:2022pah}. Figure \ref{kick_estimates_model_comparison} explains why the retention rates in Figure \ref{retention_rates} are slightly larger when using \texttt{NRSur7dq4Remnant}. Having smaller kick velocities means a higher chance of black holes being retained. 

In codes that employ \texttt{PRECESSION}, the retention rate of 1G+1G mergers will not be strongly affected by kick biases, as black holes in these systems are assumed to have negligible spin, and their kick estimates are not expected to be strongly biased. However, for binaries involving second- or higher-generation black holes (e.g., 1G+2G or 2G+2G), where black holes have non-zero spins, kick estimates will likely be biased (as shown by Figure \ref{kick_estimates_model_comparison}), leading to corresponding biases in retention rates (Figure \ref{retention_rates}) and therefore, the spin distribution of retained black holes. For this reason, we adopt a more accurate fitting formula, \texttt{NRSur7dq4Remnant}, for our study.

\section{Spin distribution in hierarchical mergers}

\subsection{1G+1G mergers}
\begin{figure}[tb]
    \centering
	\includegraphics[width=0.45\textwidth]{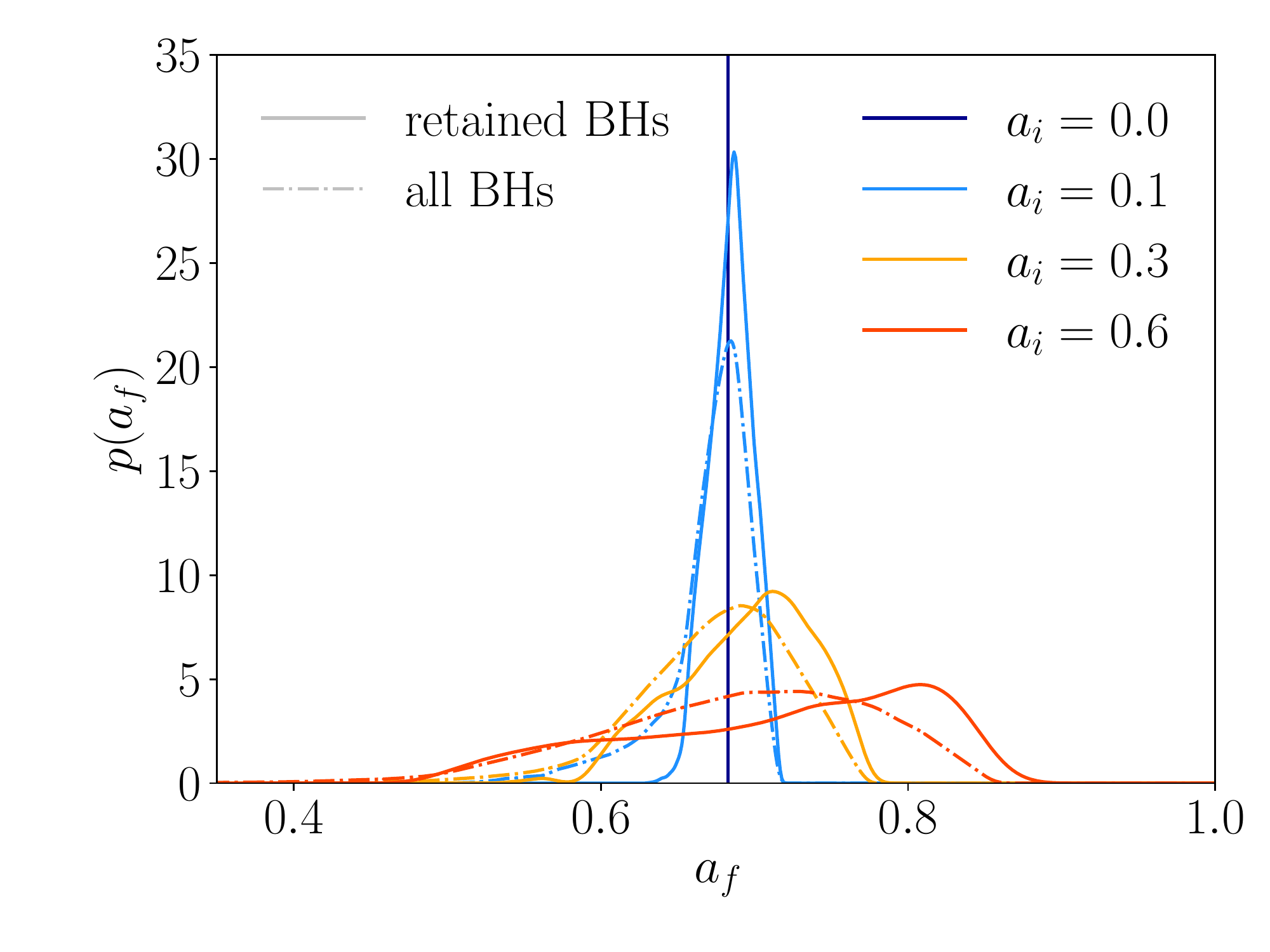}
	\caption{Spin distribution of 1G+1G merger remnants considering different initial spin magnitudes. We assume isotropic spin orientations, and consider the escape velocity profile and the mass ratio distribution of 1G+1G mergers from the CMC data. The dashed lines indicate the spin distributions of all remnants, while solid lines indicate the spin distributions of the retained ones. For the $a_i = 0.0$ case, the vertical line is centered at $a_f = 0.69$. The plot shows that the spin distribution of the retained black holes shifts towards values larger than $a_f = 0.69$ as we increase the spin magnitude of the 1G black holes.}
	\label{spin_distribution_1G1G}
\end{figure}

As we mentioned earlier, the final spin is determined by the initial binary properties. When restricting to retained black holes, we consider only those binaries for which the kick velocity is lower than the escape velocity. The kick velocity, which is determined by the same parameters as the final spin, selects which combinations of mass ratio and initial spins contribute to the spin distribution.

In Figure \ref{spin_distribution_1G1G}, we show the spin distribution of 1G+1G merger remnants, assuming different initial spin magnitudes. We use the mass ratio distribution of 1G+1G mergers and the escape velocity profile from the CMC catalog data, averaged over cluster properties. The solid lines in the figure show the spin distribution of the retained black holes, while the dashed lines show the distribution of all remnants. 

As expected, the spin distribution of all remnants, including the ejected ones, is centered around $a = 0.69$. This is consistent with earlier studies \citep[e.g.][]{Fishbach:2017dwv}. However, when looking at the spins of the retained black holes, we find that the distribution peaks at slightly higher values compared to the distribution of all remnants. The difference between the two distributions becomes more apparent as we increase the initial spin magnitude and is clearly visible for spin magnitudes of $a_i \gtrsim 0.3$. In fact, for initial spins of $a_i = 0.6$ the distribution peaks at $a_f \sim 0.8$. In addition, as we increase the initial spin magnitude, the distribution becomes wider, having more support for larger spins.

At the same time, as we increase the birth spin magnitudes, the mergers of these black holes produce larger kick velocities, and therefore, fewer remnants are retained in the cluster (see Figure~\ref{retention_rates}). Therefore, the chances of these remnants merging again reduce as we increase the initial spin magnitude.

\begin{figure}[tb]
        \centering
	\includegraphics[width=0.45\textwidth]{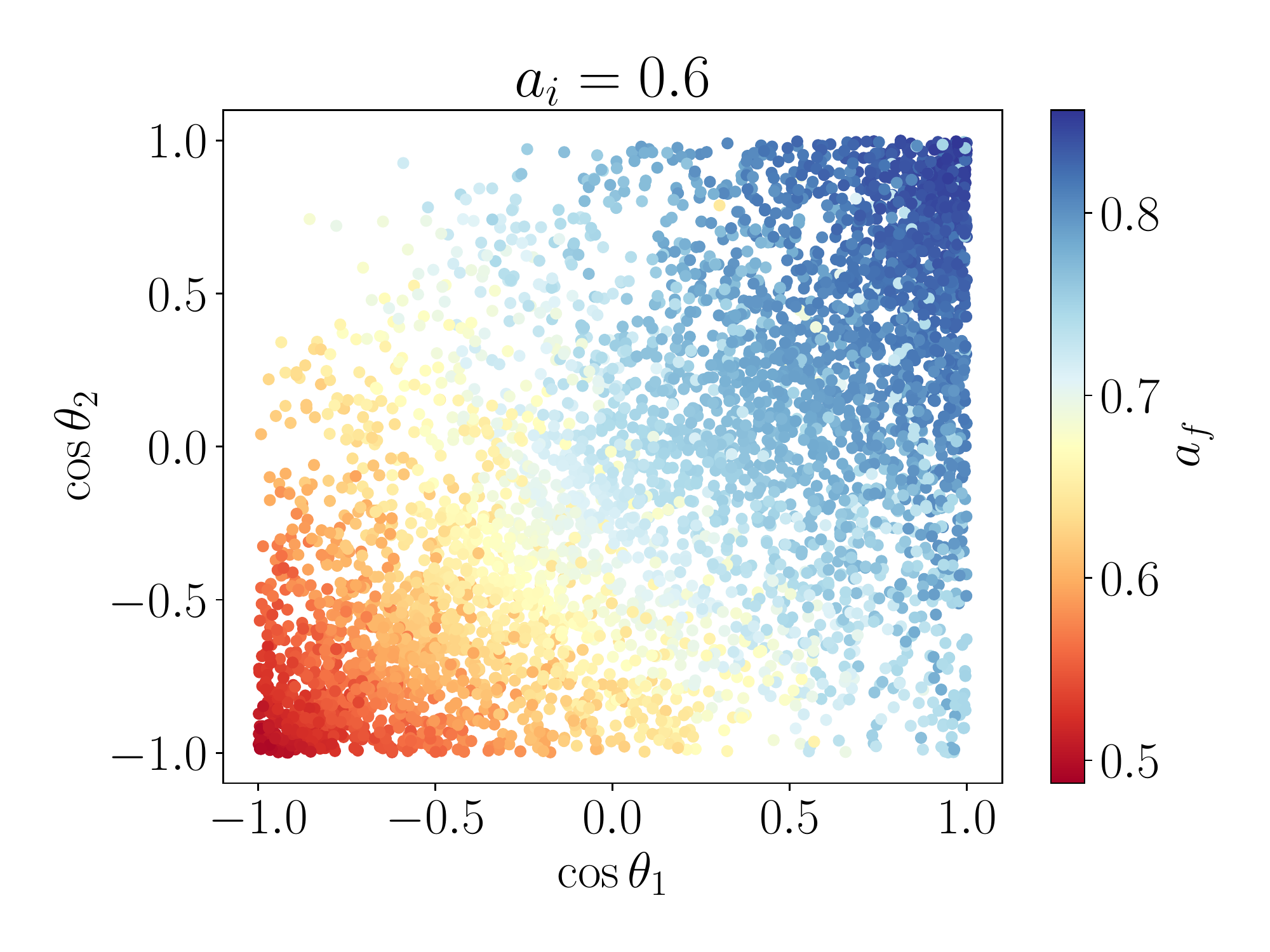}
	\caption{Spin tilts of the retained remnants' parents, considering that the initial spin magnitudes are $a_i = 0.6$. The points are colored according to the final spin value, as indicated by the colormap.}
	\label{parent_spin_orientations_1G1G}
\end{figure}

% understanding why the distribution peaks at higher values
These results show that if birth spins are non-zero, retained black holes have spins larger than $a_f = 0.69$. The exact spin value depends on the birth spin magnitude and other parameters, as we will see later in this section. To understand why the spin distribution peaks at higher values than 0.69, we investigate whether the binaries that produce retained remnants have any common properties. First, we look at the initial spin orientations of these binaries. Figure \ref{parent_spin_orientations_1G1G} shows the initial spin tilts for the case where the initial spin magnitudes are $a_i = 0.6$, as this is the case where the distinction between the spin distributions of all remnants and that of the retained remnants is the most visible (see Figure \ref{spin_distribution_1G1G}). In this plot, each dot represents one binary merger. The blank region in the upper left corner of the plot tells us that binaries with a primary spin anti-aligned with the orbital angular momentum and a secondary spin aligned produce black holes that are generally ejected. We find that 43\% of the retained remnants have both parent spins above the orbital plane, supporting final spins of $a_f > 0.75$. 30\% of retained remnants have parents with opposite spin tilts and close to the orbital plane, leading to final spins between $a_f \in (0.65, 0.75)$, while 27\% of retained remnants have both parent spins below the orbital plane producing final spins of $a_f < 0.65$. This explains why the spin distribution of retained black holes peaks at spin values larger than 0.69. 

As we mentioned previously, spin-aligned binaries have significantly lower kick velocities than binaries with misaligned spins (see Figure \ref{kicks_from_double-spin_binaries}). This explains why clusters prefer to retain binaries with close-to-aligned spins. At the same time, because the spins are close to aligned with the orbital angular momentum of the system, these spin orientations will lead to larger final spins than 0.69, which explains why the spin distribution peaks at final spin values larger than 0.69. 

The other parameter that influences the kick velocity is the mass ratio. As shown in Figure \ref{mass_ratio_distributions}, the mass ratio distribution of 1G+1G mergers peaks at $q = 1$ and has a tail that decreases to zero around $q \sim 2.5$. In binaries with equal masses, aligned and anti-aligned spin configurations lead to similar kick velocities (see Figure \ref{kicks_from_double-spin_binaries}). However, in binaries with $q > 1$, aligned-spin systems have smaller kicks than those with anti-aligned spins. Therefore, for unequal-mass 1G+1G mergers, we expect to find a stronger preference to retain remnants from aligned-spin binaries than any other spin configuration. 

\begin{figure}[tb]
        \centering
        \subfloat{%
		\includegraphics[width=0.9\linewidth]{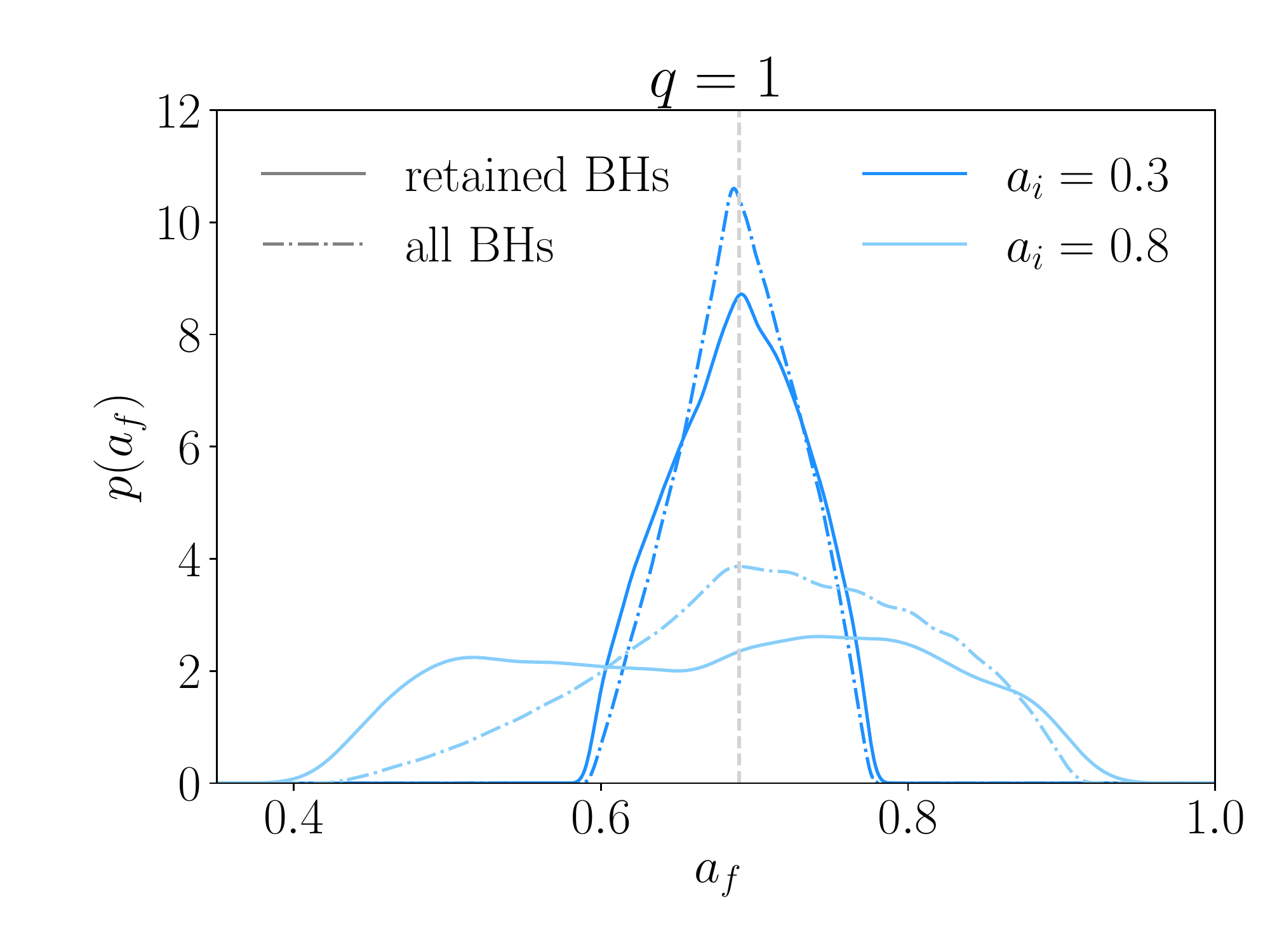}}
	\vspace{0.001cm}
        \subfloat{%
		\includegraphics[width=0.9\linewidth]{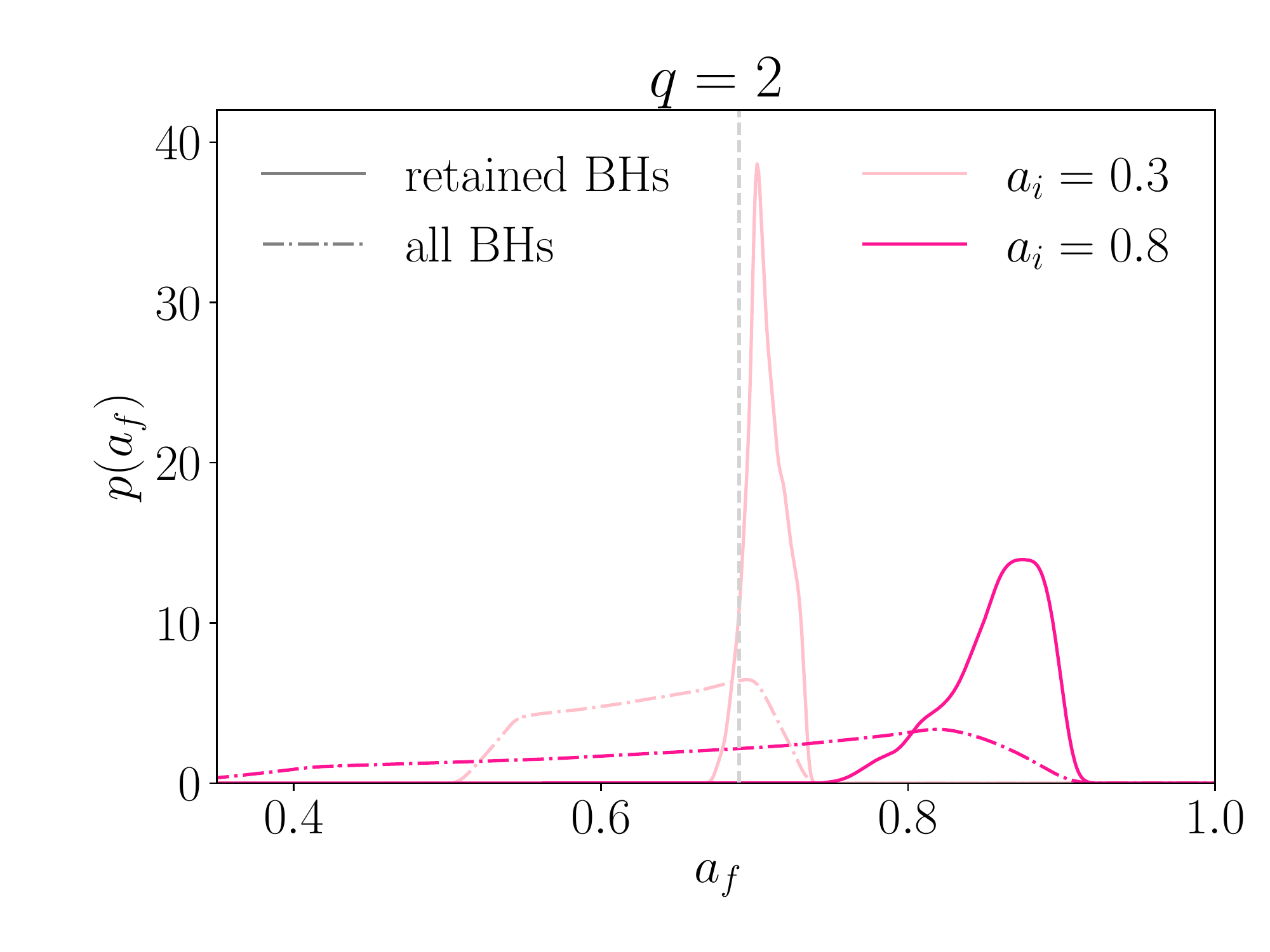}}
	\caption{Spin distribution of 1G+1G mergers, assuming that the mass ratio has a fixed value and non-zero initial spin magnitudes. The top plot shows the spin distribution considering binaries with $q = 1$, while the bottom plot shows the spin distributions assuming $q =2$. Solid lines indicate the spin distribution of retained remnants, while dashed curves indicate the distribution of all remnants, including the ejected ones. The grey vertical line is centered at $a_f = 0.69$ and is included as a reference. We assume initial spins have random isotropic orientations.}
	\label{impact_of_mass_ratio_1g1g}
\end{figure}

To further illustrate the impact of the initial binary properties on the final spin distribution, we repeat the calculation assuming that all 1G+1G mergers have a mass ratio with a fixed value. First, we test $q = 1$, and then $q = 2$. Figure \ref{impact_of_mass_ratio_1g1g} shows the spin distribution of these two cases assuming initial spins of $a_i = 0.3$ and $a_i = 0.8$. When one assumes that all 1G+1G mergers have a mass ratio of $q = 1$, the spin distribution of the retained remnants is centered at $a_f \sim 0.7$ independent of the initial spin magnitude. However, when we assume that binaries have a mass ratio value of $q = 2$, the spin distribution of the retained remnants peaks at values slightly higher than $a_f = 0.7$ in the case of initial spins of $a_i = 0.3$, and above $a_f = 0.8$ for initial spins of $a_i = 0.8$. We note that the distributions in Figure \ref{impact_of_mass_ratio_1g1g} do not represent the expected spin distribution, as the mass ratio that we used is not the expected mass ratio distribution of 1G+1G mergers in globular clusters. However, the purpose of including these results is to show that the final spin distribution of retained black holes is highly sensitive to the mass ratio and the initial spins of the binaries. 

\subsection{Impact of a cluster's escape velocity}\label{sec:impact_of_escape_velocity}
\begin{figure}[tb]
        \centering
	\includegraphics[width=0.47\textwidth]{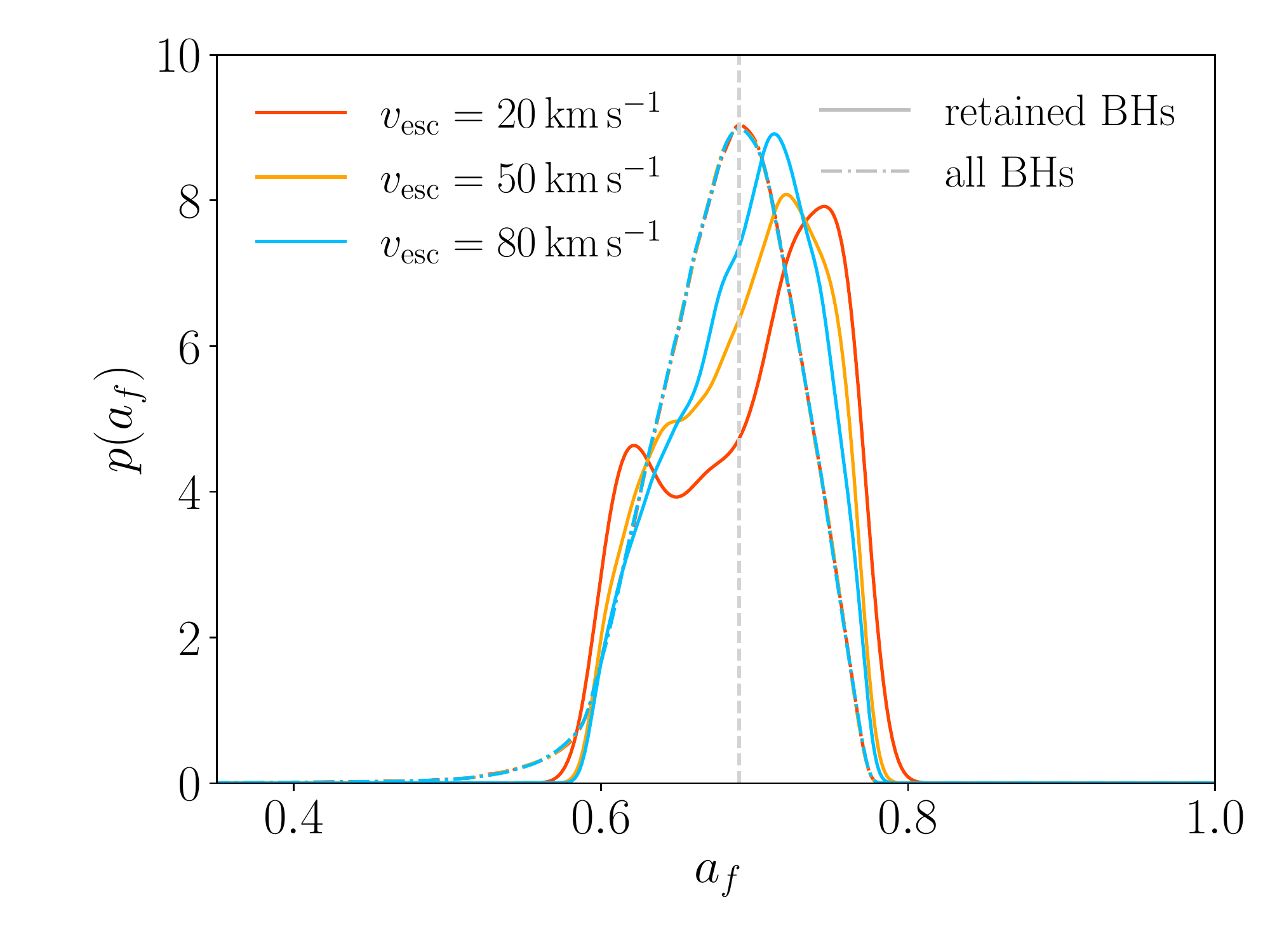}
	\caption{Spin distribution of 1G+1G merger remnants, assuming birth spins of $a_i = 0.3$ and a fixed value of the escape velocity indicated by the colors in the legend. Solid lines show the distribution of retained black holes, while dashed lines show that of all black holes. The dashed vertical line indicates $a_f = 0.69$ as a reference.}
	\label{impact_escape_velocity_1G1G}
\end{figure}

The question of whether a black hole is retained or not depends on the kick velocity but also on the escape velocity. So far, we have considered the escape velocity profile typically found in globular clusters. More massive clusters, such as nuclear star clusters, have higher escape velocities, meaning that they are able to retain more black holes. As we increase the escape velocity of a cluster, the spin distribution of retained black holes becomes more similar to the distribution of all remnants. Figure \ref{impact_escape_velocity_1G1G} shows this effect assuming initial spins of $a_i = 0.3$ and a fixed value for the escape velocity. In the plot, different colors indicate different escape velocities. If the escape velocity decreases, the peak of the spin distribution shifts towards larger spin values. 

A notable feature in the spin distribution is the bimodality that emerges, particularly in the case with $v_{\textbf{esc}} = 20\,{\rm km\,s^{-1}}$. Binaries resulting in final spins of $a_f \sim 0.7$ are suppressed in the distribution because they generally have in-plane spins and therefore receive larger kicks than aligned or anti-aligned spins (see Figure \ref{kicks_from_double-spin_binaries}). In contrast, binaries with aligned and anti-aligned spin, which are more likely to be retained, lead to slightly higher and lower final spins. Because the spin distribution depends on the escape velocity, we can expect clusters with different properties to have slightly different spin distributions.

As we lower the escape velocity to $v_{\textbf{esc}} = 20\,{\rm km\,s^{-1}}$, we must carefully account for the systematic errors in \texttt{NRSur7dq4Remnant}. The uncertainty in the kick estimate may prevent us from conclusively determining whether the kick is greater or less than the escape velocity, raising the question of whether the observed bimodality is physical. For generic spin configurations, the kick estimate has an average $1\sigma$ error of approximately $15\,{\rm km\,s^{-1}}$. However, for configurations where the kicks are smaller than $v_{\textbf{esc}} = 20\,{\rm km\,s^{-1}}$, the average error is reduced to about $10\,{\rm km\,s^{-1}}$.

\begin{figure}[tb]
        \centering
	\includegraphics[width=0.47\textwidth]{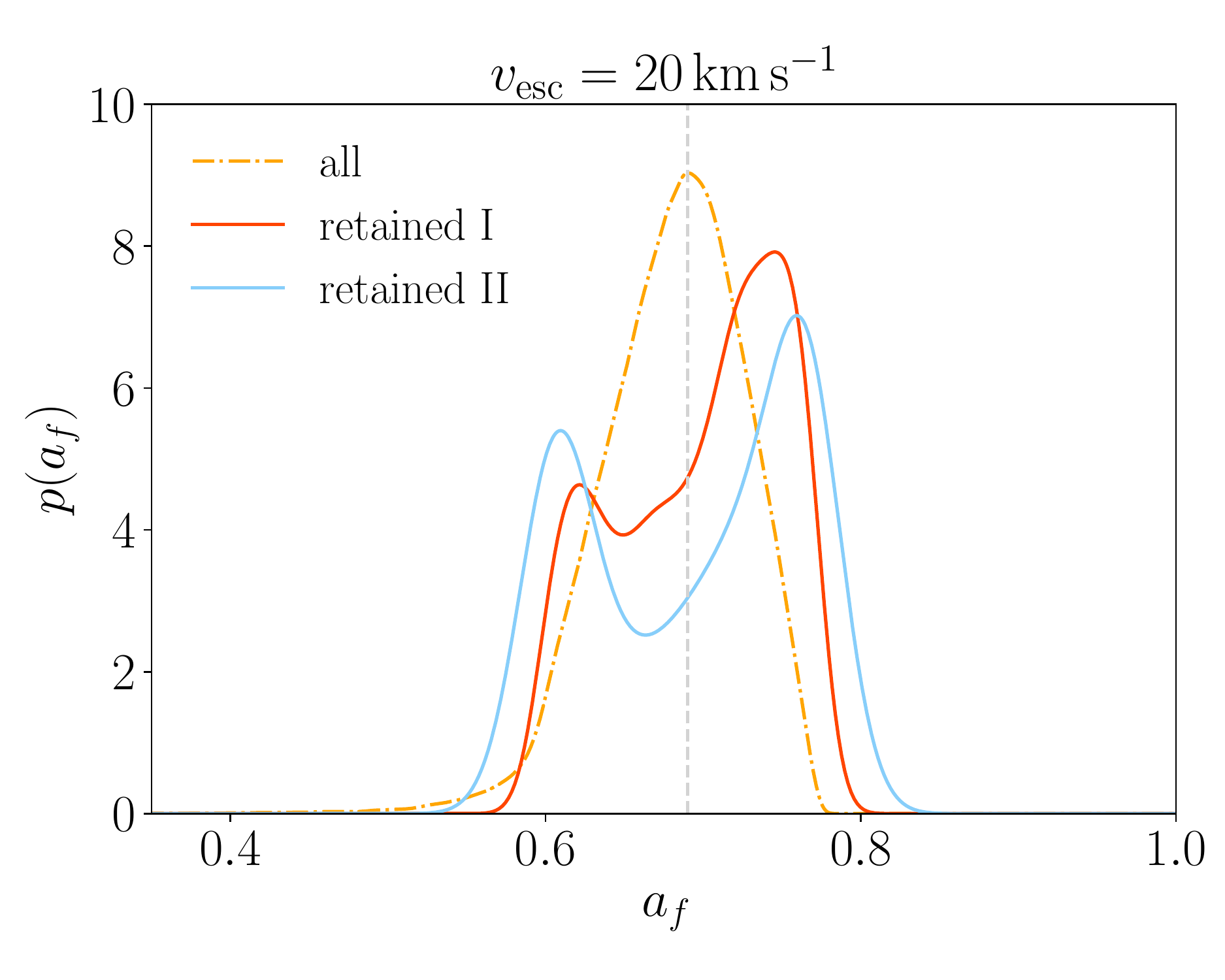}
	\caption{Spin distribution of 1G+1G merger remnants, assuming birth spins of $a_i = 0.3$ and an escape velocity of $v_{\textbf{esc}} = 20\,{\rm km\,s^{-1}}$. The orange curve shows the spin distribution of all remnants, the red curve represents the distribution of retained remnants, and the blue curve represents the spin distribution of a subset of the retained remnants for which the kick estimate, including its $1\sigma$ error, remains below the escape velocity. The dashed vertical line indicates $a_f = 0.69$ as a reference.} 
	\label{bimodality_uncertainty}
\end{figure}

To assess the uncertainty in the distribution, we compute the spin distribution considering a subset of retained remnants for which the kick velocity is well resolved and below the escape velocity. Specifically, we include only those cases where the kick plus its $1\sigma$ error remains below the escape velocity. This distribution only includes 15\% of the retained remnants. Figure \ref{bimodality_uncertainty} compares our original distribution (in red), which includes remnants for which the kick may not be well resolved and could exceed the escape velocity within its uncertainty, with our new distribution (in blue), which includes only remnants for which the kick estimate confidently remains below the escape velocity. The differences between the two distributions reflect the uncertainty in the relative heights of the two peaks, yet the bimodality remains present.

Given kick uncertainties, the precise escape velocities at which the bimodality emerges cannot be well constrained. This may be resolved in the future once a more accurate fitting formula is available.

In contrast, clusters with large escape velocities can retain black holes more efficiently and therefore, they are able to host multiple generations of mergers more easily. In the following, we investigate what is the spin distribution over several generations.

\subsection{Does the spin distribution converge after several mergers?} \label{subsection_on_convergence}

\begin{figure*}[tb]
    \centering
    \subfloat{%
        \includegraphics[width=0.45\linewidth]{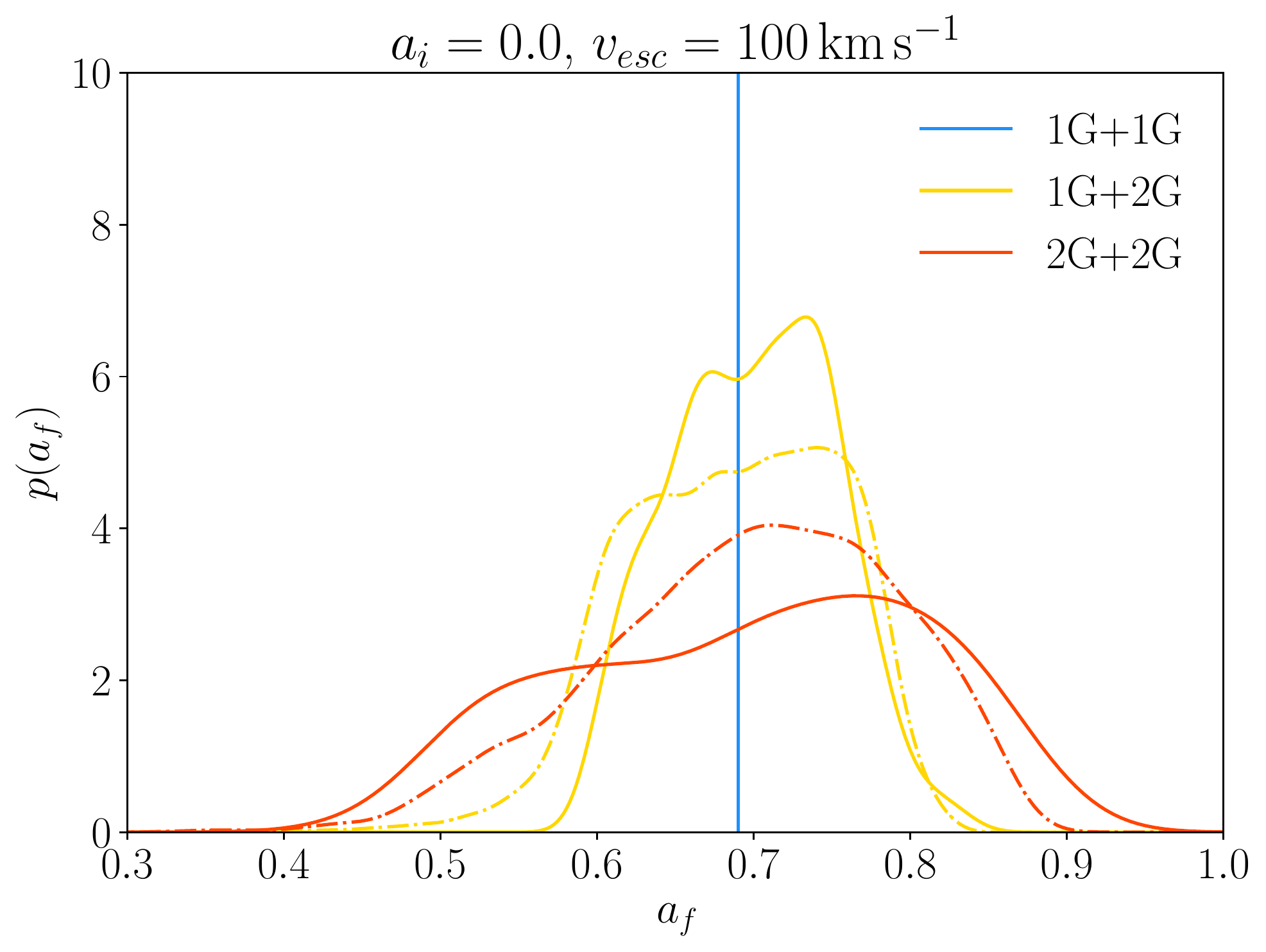}}
    \subfloat{%
	\includegraphics[width=0.45\linewidth]{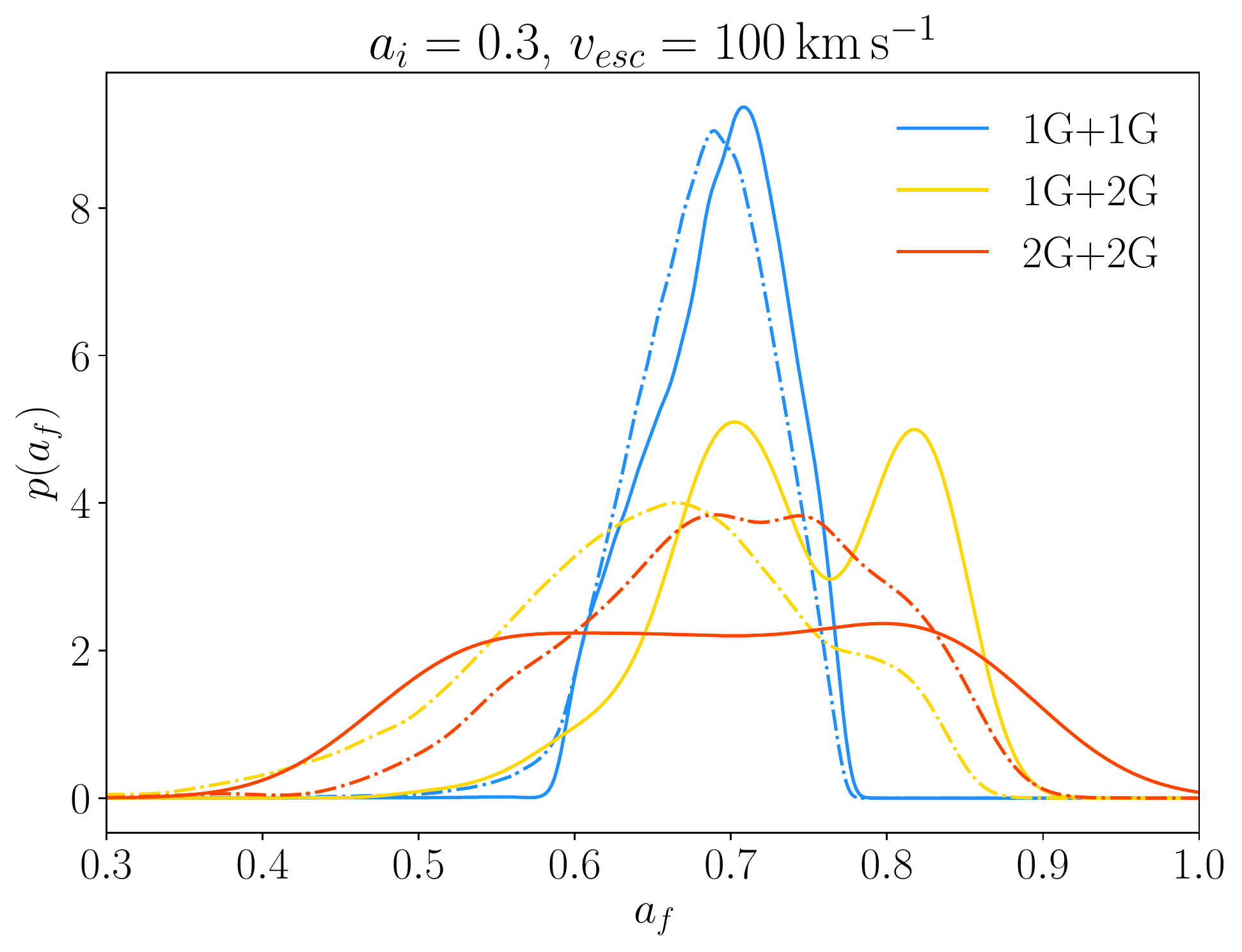}}
    
    \subfloat{%
        \includegraphics[width=0.45\linewidth]{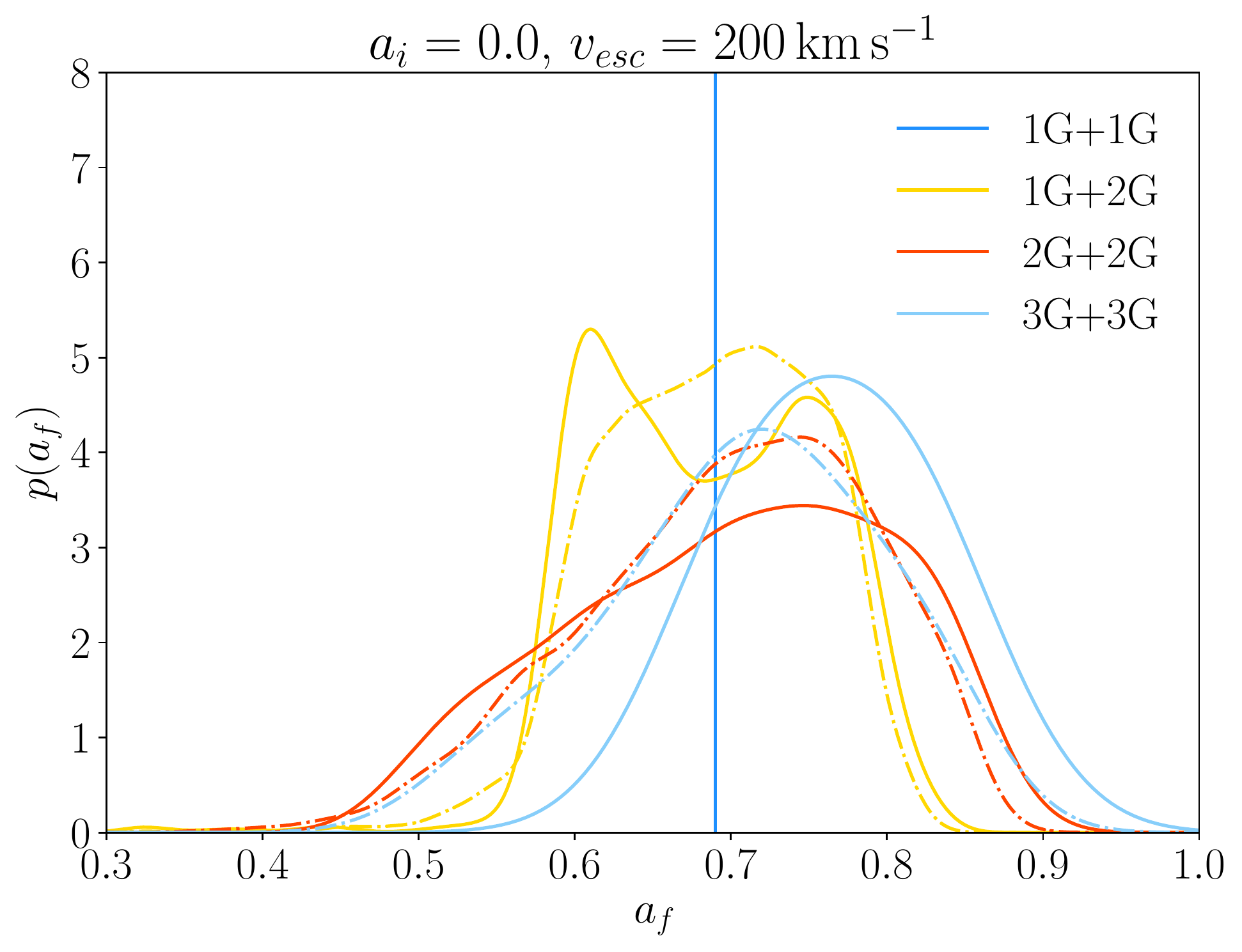}}
    \subfloat{%
        \includegraphics[width=0.45\linewidth]{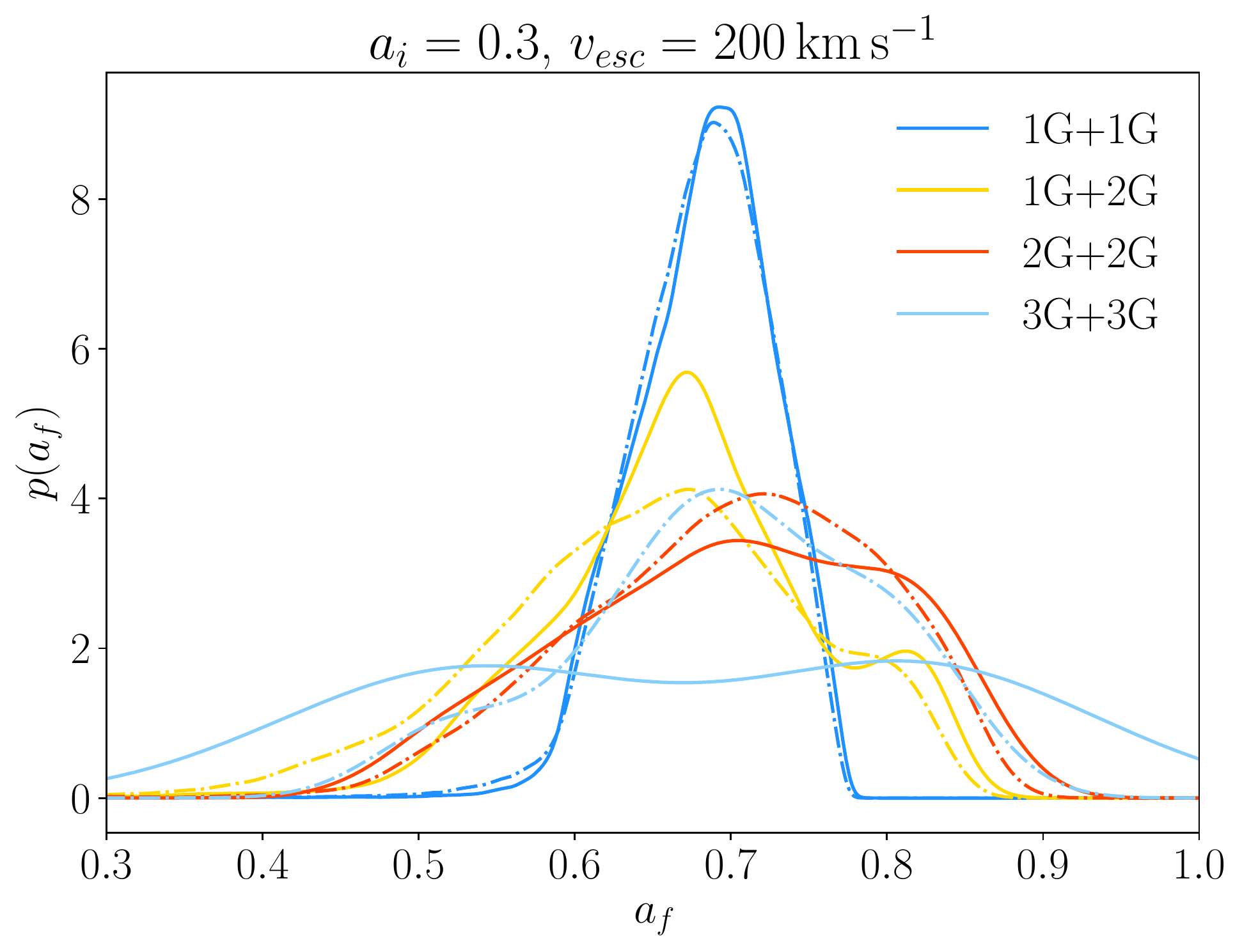}}
    
	\caption{Panels show the spin distribution of different merger generations assuming different initial spins and escape velocities. The panels in the first column assume an initial spin of $a_i = 0.0$, while the panels in the second column assume $a_i = 0.3$.  The panels in the first row assume an escape velocity of $v_{\text{esc}} = 100\,{\rm km\,s^{-1}}$, while the panels in the second row assume $v_{\text{esc}} = 200\,{\rm km\,s^{-1}}$. The spin distribution of retained remnants is shown in solid lines, while the spin distribution of all remnants is shown in dashed lines. In the case of 1G+1G mergers with $a_i = 0.0$, all remnants are retained and have spins of $a_f = 0.69$, which we indicate with a vertical line.}
	\label{convergence}
\end{figure*}

As shown in the previous section, given the mass ratio, initial spins and escape velocity, first-generation black-hole binaries produce remnants with a unique spin distribution. We now investigate whether the spin distribution of retained black holes converges after several generations. Figure \ref{convergence} summarises our results. 

We find that the spin distribution of \textit{retained} remnants is different for different black-hole generations and does not converge to a unique distribution. Different initial conditions lead to significantly different distributions of the highest-generation mergers retained, namely, 2G+2G or 3G+3G binaries. In some cases, the distribution flattens out after several mergers, as shown by the panels in the second column of Figure \ref{convergence}, supporting a wide range of spin values, while in other cases, it peaks around $a_f \sim 0.8$, as shown by the panels in the bottom left. Our results show that a $n$-th generation black hole that is retained does not necessarily have a spin of $a_f \sim 0.7$, but its value will range between $a_f \in (0.4, 1)$ depending on the spin magnitudes and orientations of the first-generation black holes, the escape velocity of its environment, and the merger generations of its parent black holes. 

At the same time, the retention rate of these mergers decreases as we consider mergers involving higher-generation black holes. Since these black holes have larger spins than first-generation black holes, they tend to produce larger kick velocities, leading to more remnants being ejected from the cluster. As a result, the likelihood of observing a binary containing a black hole from a higher-generation merger is lower than for 1G+1G mergers. Only highly massive star clusters, such as nuclear star clusters, have sufficiently high escape speeds to retain black-hole remnants with large kicks, allowing multiple generations of mergers.

In contrast, the spin distribution of \textit{all} remnants converges to a near-universal distribution after two merger generations, regardless of the initial spin magnitude, the escape velocity and the black-hole generations. Figure \ref{convergence} shows how the spin distribution of 2G+2G (red) and 3G+3G mergers (light blue) is almost identical in the two lower panels. These distributions are consistent with the findings of \cite{Fishbach:2017dwv}.

\section{What if some 1G+1G binaries have near-aligned spins?}\label{sec:as_binaries}

Although spin alignment mechanisms are usually associated with isolated binary evolution channels, about 10\% of the black holes produced in dense star clusters may also have aligned spins \citep{Kiroglu:2025bbp}. In this section, we allow some fraction of the 1G+1G binaries to have near-aligned spins and investigate whether this can impact the distribution of the final spin. As aligned-spin binaries have significantly smaller kicks than isotropic-spin binaries, we can expect the majority of the near-aligned spin binaries to be retained and contribute to the spin distribution of black holes in a cluster.

To investigate this question, we build a new toy model in which we randomly pair black holes. In this model, the majority of 1G+1G binaries have random, isotropic spins, except for a fraction that has near-aligned spins. We test three different fraction values: 25\%, 50\% and 75\%. In the case of the near-aligned spin binaries, the spin tilt of each black hole is drawn from a Gaussian distribution in the cosine of the tilt angle, $\cos\theta_i$, centered at $\cos(\ang{0})$ and with a standard deviation of $\sigma = \cos(\ang{10})$, while the spin azimuthal angle is drawn from a uniform distribution in $\phi \in [0, 2\pi]$. Regarding the spin magnitude, we test two different values: $a_i = 0.1$ and $a_i = 0.3$. In the same way as in previous sections, we use the mass ratio distribution of 1G+1G and 1G+2G mergers form the CMC data, while for higher generation mergers, we use the fixed value $m/n$, where $m$ is the generation of the primary black hole and $n$ the generation of the secondary. We use a fixed value of the escape velocity $v_{\text{esc}} = 50\,{\rm km\,s^{-1}}$, which is in the range of the typical escape velocities in globular clusters. 

\begin{figure*}[tb]
        \centering
         \subfloat{%
		\includegraphics[width=0.45\linewidth]{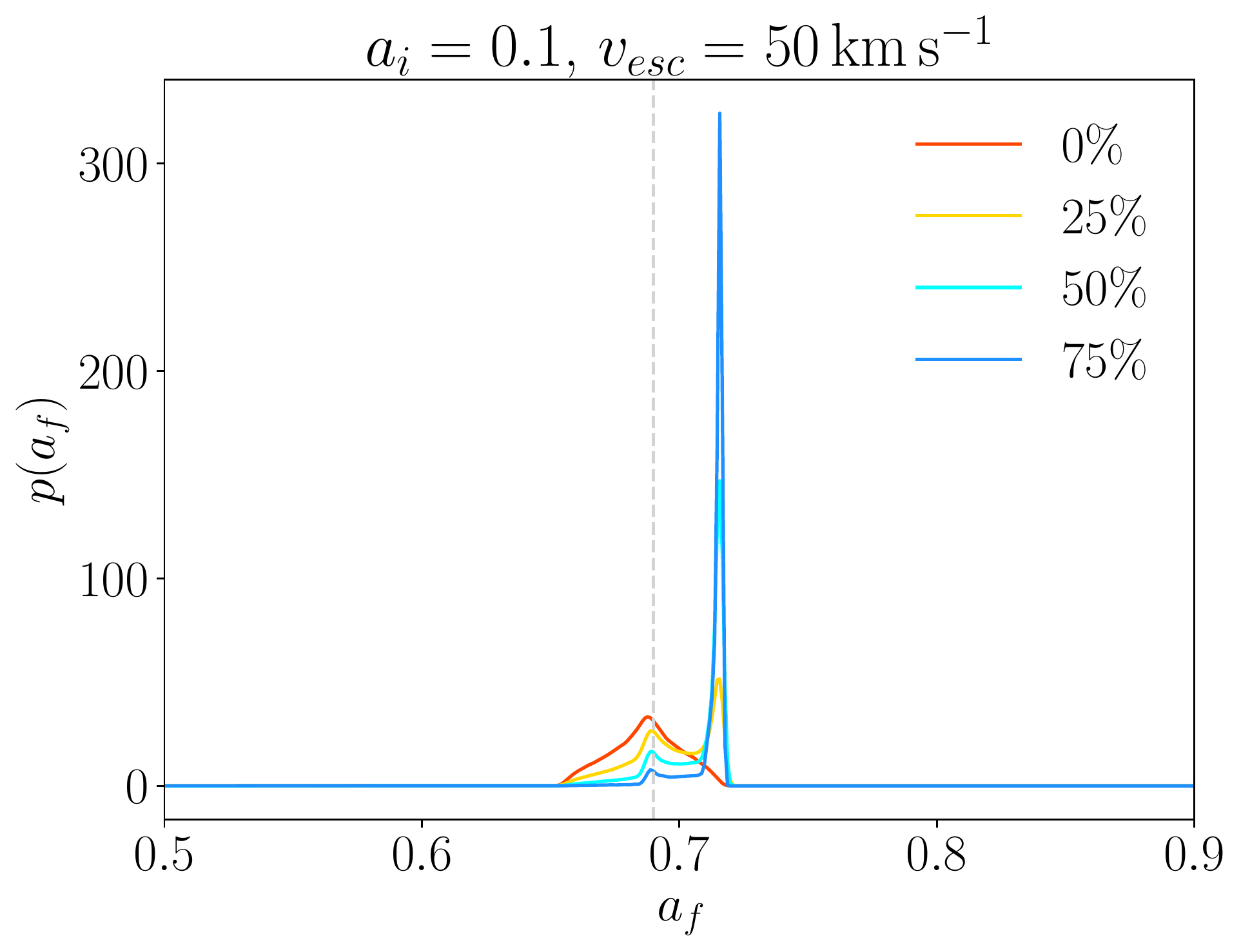}} 
        \subfloat{%
		\includegraphics[width=0.45\linewidth]{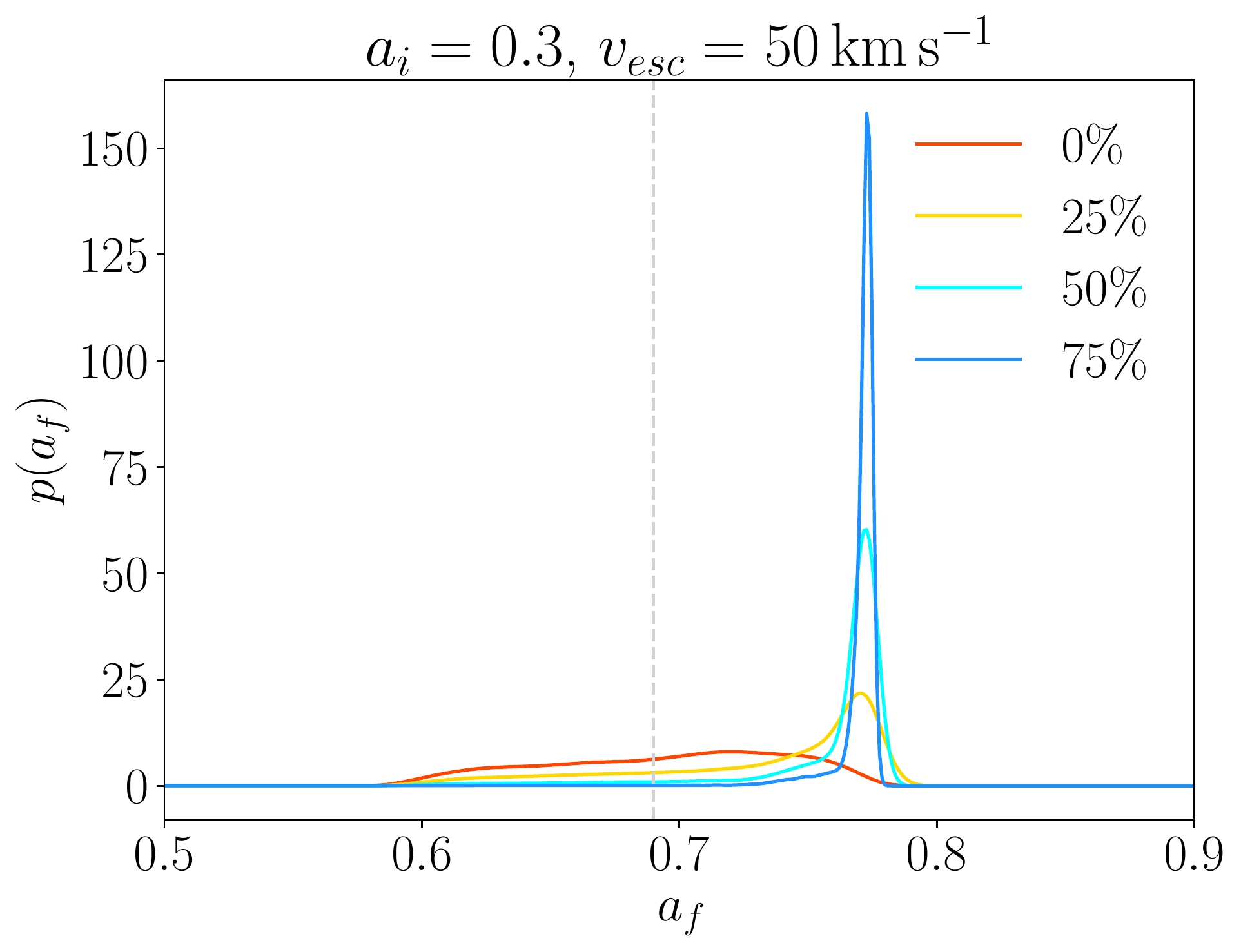}}
	\caption{Spin distribution of 1G+1G mergers, assuming different fractions of near-aligned spin binaries: 0$\%$ (red), 25$\%$ (yellow), 50$\%$ (turquoise) and 75$\%$ (blue). The left panel considers an initial spin magnitude of $a_i = 0.1$, while the right panel considers spins of $a_i = 0.3$. The escape velocity is fixed to $v_{\text{esc}} = 50\,{\rm km\,s^{-1}}$. For reference, the dashed vertical line indicates the final spin value at $a_f = 0.69$.} 
	\label{impact_of_AS_fraction_on_1g1g}
\end{figure*}

Figure \ref{impact_of_AS_fraction_on_1g1g} shows the spin distribution of 1G+1G mergers assuming different fractions of near-aligned spin binaries for an escape velocity of $v_{\text{esc}} = 50\,{\rm km\,s^{-1}}$. Here, the left panel assumes a spin magnitude of $a_i = 0.1$, while the right panel assumes $a_i = 0.3$. As shown by the left panel, when the fraction of binaries with near-aligned spins is 25\%, the final spin distribution has two peaks. The first peak is centered at $a_f = 0.69$, while the second one peaks at $a_f = 0.72$. Binaries with aligned spins lead to larger final spin values than other initial spin orientations. In our case, the presence of aligned spin binaries produces a second peak in the spin distribution. As we increase the fraction of near-aligned spin binaries, the first peak loses support, while the second peak becomes more dominant. 

As expected, larger initial spin magnitudes lead to larger final spin values. When we increase the initial spin magnitude to $a_i = 0.3$ (see right panel), the second peak is centered at higher final spin values, in particular, at $a_f = 0.77$. Larger initial spins also mean larger kick velocities on average, causing fewer binaries to be retained. Between the aligned spin binaries and the isotropic spin binaries, the second ones are more susceptible to being ejected than the first, as they generally produce larger kicks (see Figure \ref{mass_ratio_dependency_of_the_kick}). This explains why, in the right panel, the spin distribution has less support around $a_f = 0.69$ compared to the case with initial spins of $a_i = 0.1$ in the left panel. Increasing the escape velocity allows the cluster to retain more binaries, increasing the support of spins around $a_f = 0.69$.

% paragraph on convergence
Regarding the spins of higher-generation merger remnants, similar to our findings in Section \ref{subsection_on_convergence}, the distribution depends on several parameters: the initial spin magnitudes and orientations, the escape velocity of the cluster and additionally, in this case, the fraction of aligned-spin 1G+1G binaries. 

Hence, the presence of spin-aligning mechanisms in a cluster can lead to black holes with spins larger than $a_f = 0.69$. The magnitude of the black-hole birth spin, the fraction of binaries affected by spin-aligning mechanisms in a cluster and the black hole generations will determine the spin magnitude distribution of the retained black holes.

\section{Conclusions}

An open question is whether black holes from hierarchical mergers can be identified in GW data. Previous studies have shown that the spin distribution of black holes from hierarchical mergers is centered at $a_f \sim 0.7$, approximately independently of the black hole's initial spin, the binary's mass ratio and the merger generation \citep{Fishbach:2017dwv, Berti:2008af}. However, the anisotropic emission of GWs imparts kick velocities to the merger remnants, which may exceed the escape velocities of many star clusters. As a result, not all merger remnants are retained within their host environments. Only those that remain bound can later form binaries, undergo mergers, and produce next-generation black holes. 

If we want to identify black holes from hierarchical mergers in GW observations, we need to focus on those that were retained in their clusters, formed binaries, and eventually merged. In contrast, ejected merger remnants are unlikely to form new binaries in the galactic field. In this paper, we have investigated what the spins of retained black holes are by considering the kick velocities produced in every binary merger. 

As the kick velocity depends on the mass ratio and the initial spins of the component black holes, not every binary configuration is equally likely to be retained. In particular, aligned-aligned binaries have significantly smaller kick velocities than binaries with random, isotropic spins. For this reason, clusters are more efficient in retaining binaries with spins close-to aligned with the angular momentum of the system. We demonstrate the impact of the mass ratio and the initial spins on the kick velocity and, in turn, on the retention of black hole merger remnants. As the spin of merger remnants is determined by the same parameters as the kick, we find that remnant kicks influence the spin distribution of black holes from hierarchical mergers. We show in detail the dependence of the spin distribution on the binary's mass ratio, the black hole birth spins, the escape velocity and the black hole generations.

In the case of 1G+1G mergers, when we consider a mass ratio distribution and an escape velocity profile typically found in globular clusters, we find that the spin distribution of retained black holes is skewed towards slightly larger spin values than 0.7 as we increase the magnitude of black hole birth spins (or decrease the escape speed of the environment). For small birth spins of $a_i = 0.1$, the distribution peaks at $a_f \sim 0.7$, as previously expected, but for moderate birth spins of $a_i = 0.6$, the spin distribution peaks around $a_f \sim 0.8$. 

Unlike the spin distribution of all remnants, we find that the spins of retained black holes do not converge to a unique distribution after several generation mergers. Different generations have different spin distributions, and different initial conditions lead to different distributions. In particular, the spins depend on the birth spin magnitudes, the mass ratio distribution, the escape velocity of the host environment and the black hole generations. We observe that higher-generation black holes can have spins between $a_f \in (0.4, 1)$, and not necessarily $a_f = 0.7$, because the binary configurations that lead to $a_f = 0.7$ are the ones most likely to get ejected from the cluster. In environments with low escape velocities, this leads to a bimodal spin distribution, with two peaks around $a_f \sim 0.6$ and $a_f \sim 0.8$, though their exact positions depend on the birth spin. These peaks arise from anti-aligned and aligned spin configurations, which, on average, produce significantly smaller kicks than other binary configurations.

In addition, we have investigated whether the presence of spin-aligning mechanisms in dense star clusters can impact the black hole spins. We find that aligned-spin binaries disproportionately produce retained merger remnants, so even a small fraction of nearly-aligned binaries leads to a second peak in the spin distribution that is centered at higher values. The exact location of the second peak depends on the birth spins. For instance, for black hole birth spins of $a_i = 0.3$, the second and most dominant peak is centered close to $a_f \sim 0.8$. In the case of higher-generation mergers, the spin distribution is determined by the fraction of binaries affected by spin-aligning mechanisms in a cluster, together with the birth spin magnitudes and the black hole generations. Once again, we find that higher-generation black holes can have a wide range of spin values, $a_f \in (0.4, 1)$. Although spin alignment mechanisms might be present in certain environments such as active galactic nuclei \citep[see e.g.,][]{Bogdanovic:2007hp, Mckernan:2017ssq}, our results do not straightforwardly apply to these, as the properties of black-hole binaries might not necessarily be the same as in dense star clusters \citep{Yang:2019cbr}.

We have investigated whether systematic errors in models of the remnant properties can influence these predictions. Specifically, we compare \texttt{PRECESSION}, the model commonly used in simulations of dense star clusters, with \texttt{NRSur7dq4Remnant} the state-of-the-art model. Our analysis shows that, on average, \texttt{PRECESSION} predicts kick velocities that are an order of magnitude higher than those of \texttt{NRSur7dq4Remnant}, resulting in biased retention rates.

With this study, we complement previous work on characterising the spins of black holes from hierarchical mergers, which can be useful to distinguish hierarchical-merger black holes in GW observations. 
{For example, we showed that in many scenarios, higher-generation black holes are more likely to have dimensional spin magnitudes $a \approx 0.5$ or $a \approx 0.8$ than the canonical value of $a = 0.7$. This is because the binaries that produce remnant black holes with $a_f = 0.7$ receive the largest kicks and are most likely to be ejected from the cluster (and thus unable to participate in future mergers).}
{Given current measurement uncertainty on black hole component spins, we do not expect current GW observations to be sensitive to these differences in the spin distributions, which will have a minor effect on well-measured parameters like the effective inspiral spin $\chi_\mathrm{eff}$. However, with thousands of GW observations expected in the upcoming observing run O5 ~\citep[e.g.][]{KAGRA:2013rdx, Kiendrebeogo:2023hzf}, it will be important to take into account the precise spin distributions of \emph{retained} black holes, rather than the canonical ``universal" approximation, to accurately measure the contribution of higher-generation black holes to the GW population. While the spin distribution from retained hierarchical mergers may be less universal than previously thought, the additional features in the spin distribution encode astrophysical properties such as the escape speed of the environment and black hole birth spins. Therefore, precise measurements of the black hole spin distribution with future catalogs will reveal detailed insights into their formation environments.}

\section*{Acknowledgments}
We are grateful to Aditya Vijaykumar, who first suggested looking at the kick estimates of CMC simulations. The authors are also grateful to Amanda Farah, Davide Gerosa, Shrobana Ghosh, Jannik Mielke, Lavinia Paiella, Frank Ohme and the CITA GW group for useful discussions. We thank Sharan Banagiri for useful suggestions during the LVK internal review. A.B.~thanks the CITA GW group for hospitality while part of this work was carried out and acknowledges support from the DAAD ``Forschungsstipendien für Doktorandinnen und Doktoranden" Scholarship and the Max Planck Society’s Independent Research Group Grant. C.S.Y.\ acknowledges support from the Natural Sciences and Engineering Research Council of Canada (NSERC) DIS-2022-568580. M.F. acknowledges support from the Natural Sciences and Engineering Research Council of
Canada (NSERC) under grant RGPIN-2023-05511, the University of Toronto Connaught Fund, and the Alfred P. Sloan Foundation. Computations were performed on the Holodeck cluster of the Max Planck Independent Research Group ``Binary Merger Observations and Numerical Relativity".

\begin{appendix}
\section{A closer look into the kick velocities of aligned-spin binaries}\label{sec:appendix}
In this section, we examine the peculiarities of kicks in binaries with aligned spins and unequal mass ratios. As we mentioned earlier, aligned spin binaries produce smaller kicks than anti-aligned spin binaries for mass ratios $q > 1$. Contrary to our intuition, we find that increasing the spin magnitude produces smaller kicks in these binaries. To understand the kick velocities of these systems, we investigate how the kick velocity is built up during the binary evolution. 

\begin{figure*}[tb]
        \centering
        \subfloat{%
		\includegraphics[width=0.45\linewidth]{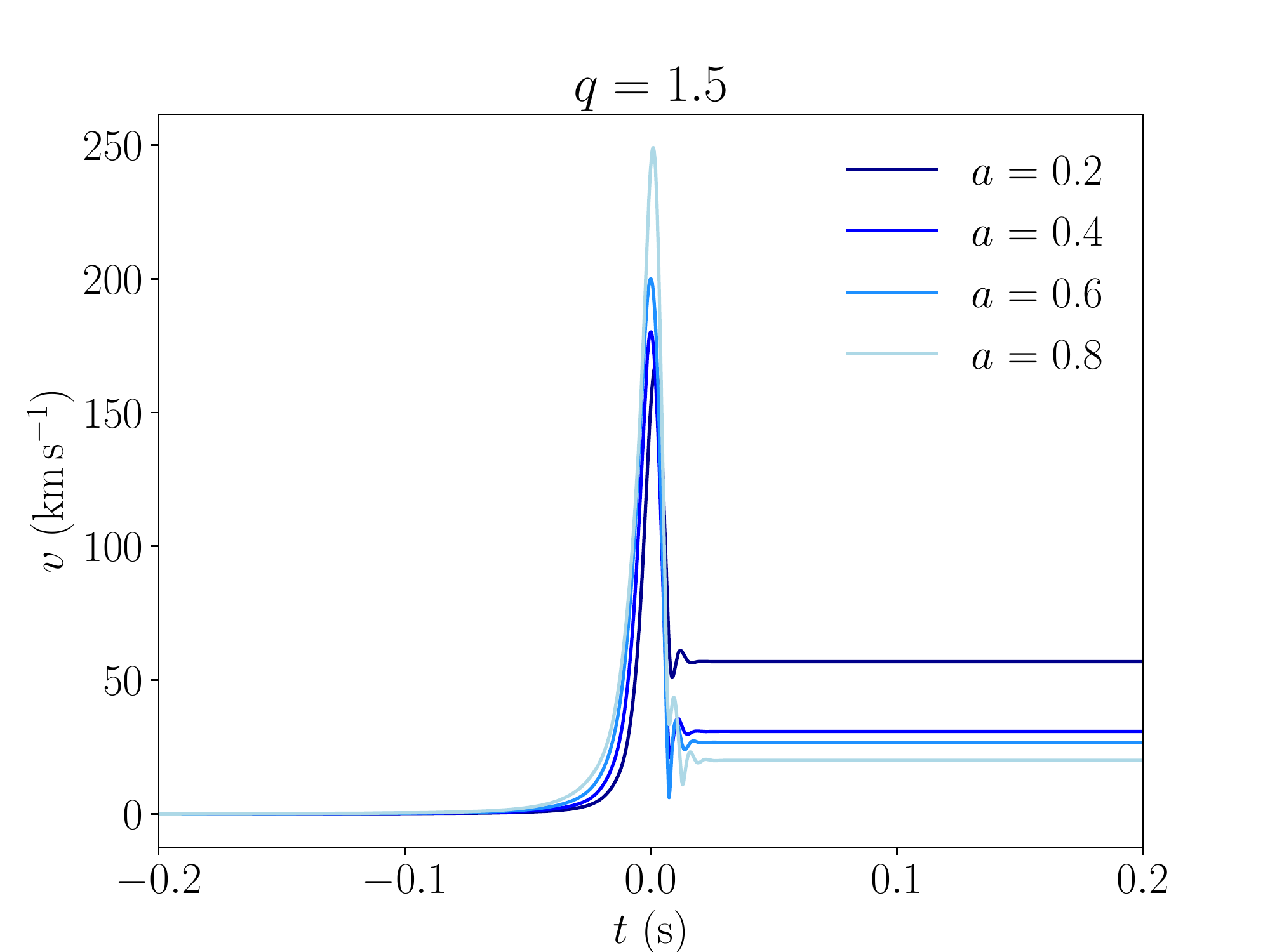}}
	\subfloat{%
		\includegraphics[width=0.45\linewidth]{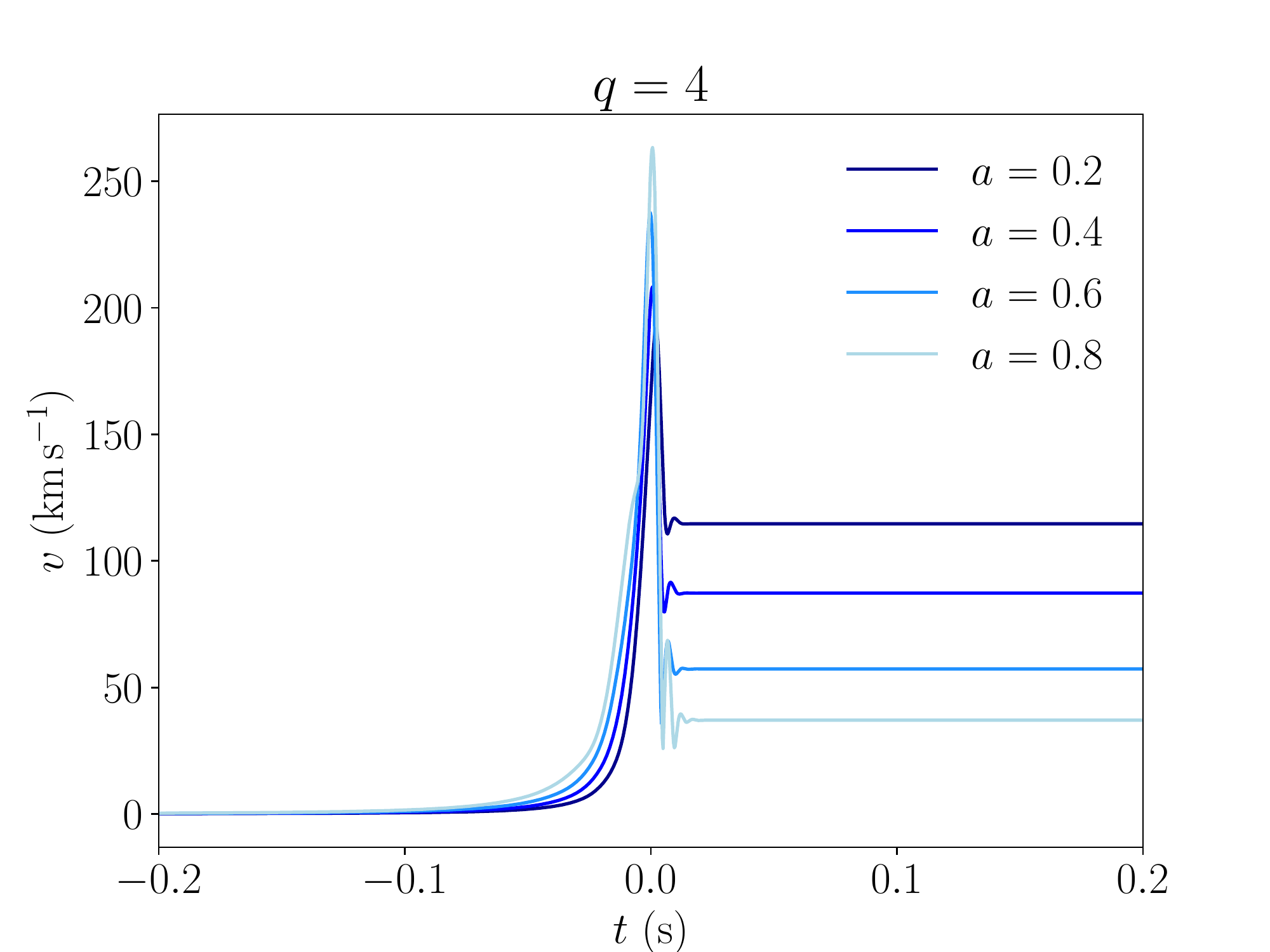}}
	\caption{Kick profile of aligned-spin binaries with spin magnitudes $a \in (0.2, 0.4, 0.6, 0.8)$, fixing the mass ratio to $q = 1.5$ and $q = 4$, respectively. We assume both black holes have the same spin magnitude. The plot shows that the maximum peak of the velocity increases with the spin magnitude. After the merger, the drop in velocity also becomes more pronounced as the spin magnitude increases. The kick profile has been computed using the waveform model \texttt{IMRPhenomXO4a}.}
	\label{kick_profile}
\end{figure*}

We calculate the velocity of the binary's center of mass as a function of time using the waveform model \texttt{IMRPhenomXO4a} \citep{Thompson:2023ase}. Figure \ref{kick_profile} shows the kick profile of aligned-spin binaries with mass ratios $q = 1.5$ (left panel) and $q = 4$ (right panel). In these plots, $t = 0\, s$ corresponds to the maximum peak of the signal amplitude, which happens at the merger. As we can see, the kick builds up in the merger, decreases drastically after $t = 0\, s$ and stabilises at a value that is much smaller than the maximum peak. The decrease in velocity has traditionally been denoted as ``anti-kick" and is thought to be related to the emission of quasi-normal modes, as the newly formed remnant is highly perturbed and radiates GWs to stabilise into a Kerr black hole. This additional emission of GWs and linear momentum flux instantaneously changes the velocity of the remnant \citep{LeTiec:2009yg, Rezzolla:2010df, Price:2011fm, Jaramillo:2011re}. This behaviour is particularly strong for aligned-aligned binaries and has previously been pointed out in the literature \citep{Healy:2014yta, Gerosa:2018qay}. 

As shown in Fig.~\ref{kick_profile}, the largest spin magnitudes have indeed the largest maximum peaks of the velocity. However, as we increase the spin magnitude, the anti-kick is stronger and the velocity of the remnant becomes smaller. This behaviour is unexpected, as one would typically expect higher spin magnitudes to produce larger kick velocities. We verified this effect with numerical relativity waveforms from the SXS Catalog \citep{Boyle:2019kee} and found the same phenomenon.

From an astrophysical point of view, this effect implies that if aligned-spin binaries are formed in a certain environment, their retention rate would increase with the initial spin magnitude, which would allow hosting several generations of hierarchical mergers. The impact of aligned-spin binaries on the spin distribution of hierarchical merger black holes is described in detail in Section \ref{sec:as_binaries}.

\end{appendix}

\bibliographystyle{aasjournal}
\bibliography{references.bib}

\end{document}